\documentclass[universe,review,accept,moreauthors,pdftex]{Definitions/mdpi} 
\firstpage{1} 
\pubvolume{1}
\issuenum{1}
\articlenumber{0}
\pubyear{2021}
\copyrightyear{2020}
\datereceived{} 
\dateaccepted{} 
\datepublished{} 
\hreflink{https://doi.org/} 

\Title{Evolution of neutron star magnetic fields}

\TitleCitation{...}

\Author{Andrei P. Igoshev$^{1}$\orcidA{}, Sergei B. Popov $^{2}$*\orcidB{} and Rainer Hollerbach $^{1}$\orcidC{}}

\AuthorNames{Andrei P. Igoshev, Sergei B. Popov and Rainer Hollerbach}

\AuthorCitation{Igoshev, A.P.; Popov, S.B.; Hollerbach, R.}

\address{%
$^{1}$ \quad Department of Applied Mathematics, University of Leeds, Leeds LS2 9JT, UK;\qquad\qquad A.Igoshev@leeds.ac.uk, R.Hollerbach@leeds.ac.uk\\
$^{2}$ \quad Sternberg Astronomical Institute, Lomonosov Moscow State University, 119234 Moscow, Russia; polar@sai.msu.ru}

\corres{Correspondence: polar@sai.msu.ru; Tel.: 
+7-495-939-5006 (S.P.)}

\abstract{
Neutron stars are natural physical laboratories allowing us to study a plethora of phenomena in extreme conditions. In particular, these compact objects can have very strong magnetic fields with non-trivial origin and evolution. In many respects its magnetic field determines the appearance of a neutron star. Thus, understanding the field properties is important for interpretation of observational data. Complementing this, observations of diverse kinds of neutron stars enable us to probe parameters of electro-dynamical processes at scales unavailable in terrestrial laboratories. 
In this review we first briefly describe theoretical models of formation and evolution of magnetic field of neutron stars, paying special attention to field decay processes. Then we present important observational results related to field properties of different types of compact objects: magnetars, cooling neutron stars, radio pulsars, sources in binary systems. After that, we discuss which observations can shed light on  obscure characteristics of neutron star magnetic fields and their behaviour. We end the review with a subjective list of open problems. 
}


\begin{document}

\section{Introduction}

Magnetic fields play a significant role in many areas of modern astrophysics, manifesting themselves in different energetic phenomena ranging from solar flares to gamma-ray and fast radio bursts. The strongest magnetic fields are found in neutron stars (NSs), where they control whether NSs are seen as normal radio pulsars, magnetars, dim isolated cooling NSs, or something else \citep{Harding2013,Perna2014AN}. Magnetic fields and neutron stars are therefore intertwined, and understanding the formation and evolution of NS magnetic fields will significantly advance our understanding of the NS population, and {\it vice versa}, systematic study of the NS population will help to understand the formation and evolution of their magnetic fields. 

Large-scale, poloidal magnetic fields control the spin-down of NSs. Typically the large-scale magnetic fields are assumed to be dipolar, as this component diminishes most slowly with distance from the star. Soon after the discovery of radio pulsars \cite{Hewish1968} it was shown that a dipole rotating in a vacuum emits electromagnetic radiation and slows down \cite{Ostriker1969}. Later \cite{Goldreich1969} it was shown  that rotation of magnetised NSs causes an electric potential to develop at the polar caps. This potential is strong enough that it can extract particles from the NS surface, and fill the NS magnetosphere with them. It was further shown that charged particles can also be created in the magnetic field by a split of high-energy photons \citep{Ruderman1975}. Recent three-dimensional numerical simulations \citep{Philippov2014} demonstrated that NS spin-down depends on the poloidal dipole field $B_p$ as follows:
\begin{equation}
P \dot P = (\kappa_0 + \kappa_1 \sin^2 \alpha) \beta B_p^2 ,
\label{e:mf}
\end{equation}
where $P$ is the star's spin period, $\dot P$ is its time derivative, $\beta$ is a constant which contains the star's moment of inertia, and $\alpha$ is the angle between the orientation of the magnetic dipole and the rotation axis (sometimes called the obliquity angle). The numerically found coefficients are $\kappa_0\approx 1$ and $\kappa_1\approx 1.2$. The same calculations show that the obliquity angle evolves slowly as:
\begin{equation}
\frac{d\alpha}{dt} = - \kappa_2 \beta \frac{B_p^2}{P^2} \sin \alpha \cos \alpha .  
\end{equation}
The numerically found coefficient $\kappa_2 \approx 1$. Radio polarisation measurements demonstrated that this angle seems to evolve on a $\sim 10^7$~year timescale \citep{Tauris1998}, leading to a situation where the orientations of the magnetic dipole and rotational axis eventually align.

\begin{figure}[h]
\includegraphics[width=10.5 cm]{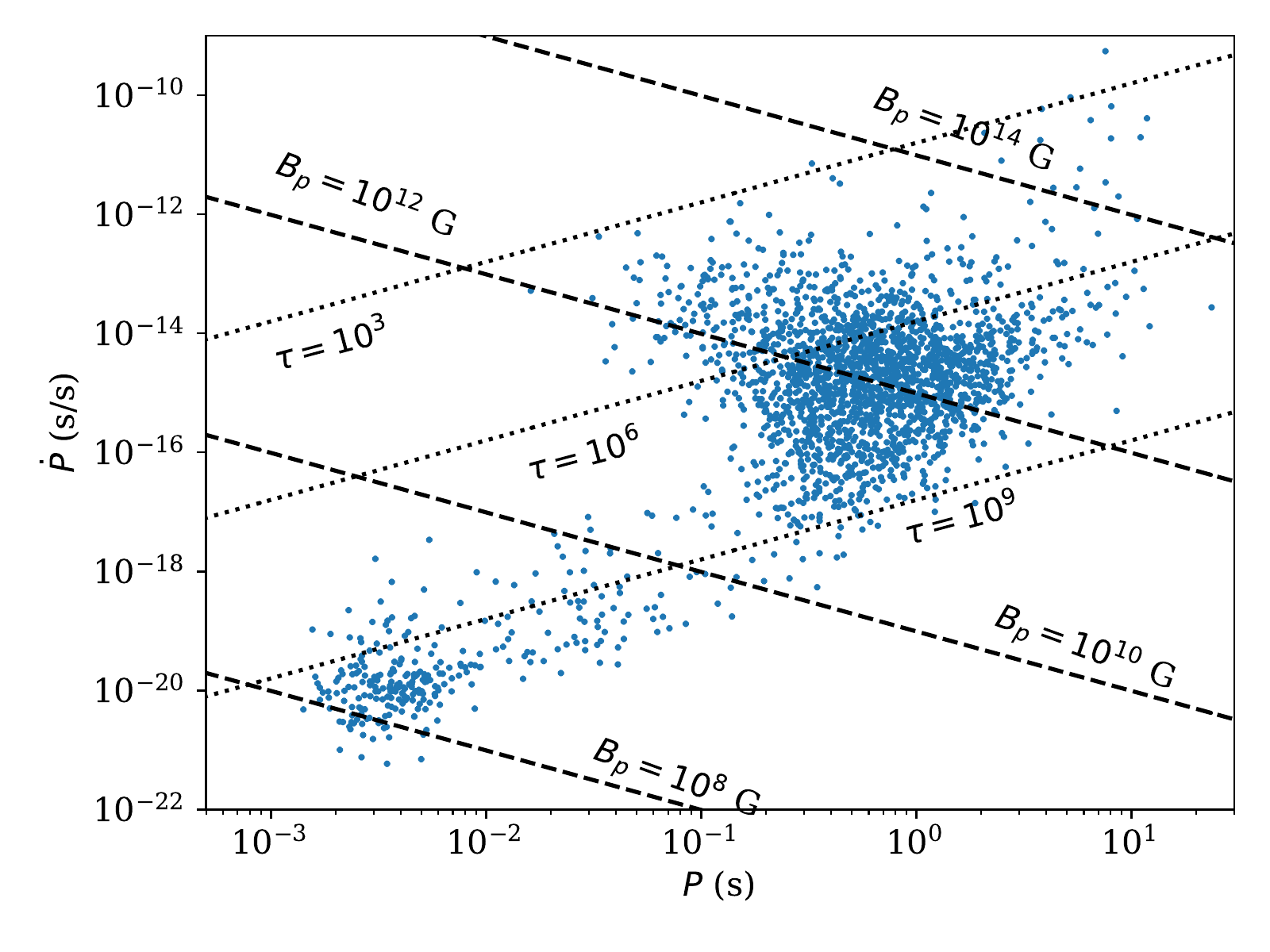}
\caption{Period~--~period derivative diagram for radio pulsars based on the ATNF pulsar database.}
\label{pdotp}
\end{figure}

Timing observations of NSs provide us with measurements of spin period $P$, its time derivative $\dot P$, and in some cases even second and third time derivatives $\ddot P$ and $\dddot P$. The instantaneous rotational period is approximated using the Taylor expansion:
\begin{equation}
P(t) = P_0 + \dot P (t - t_0) + \frac{\ddot P}{2} (t - t_0)^2 + \frac{\dddot P}{3!} (t - t_0)^3 + \ldots ,
\end{equation}
where $P_0$ is the initial spin period and $t_0$ is the epoch when period and period derivatives were measured. 
For non-accreting radio pulsars $\dot P > 0$ due to spin-down; thus knowledge of $P$ and $\dot P$ allows us to estimate $B_p$, see eq.~(\ref{e:mf}). Nevertheless, the most frequently used equation is \citep{LorimerKramer2004}:
\begin{equation}
B_p = 3.2\times 10^{19} \sqrt{P\dot P} \;\mathrm{G} .
\end{equation}
In this estimate the angle $\alpha$ is not used, because it is often not measured. Typical values of dipolar NS magnetic fields range from $10^8$~G to a few $10^{14}$~G, see Figure~\ref{pdotp}. Smaller magnetic fields $10^8-10^{10}$~G are typical for recycled radio pulsars, i.e.\ for NSs which went through an extended period of accretion in a low-mass binary. Larger magnetic fields $10^{11}-10^{13}$~G are typical for normal radio pulsars, and the strongest dipolar fields $10^{13}$~--~$10^{15}$~G are found in magnetars.

It is easy to derive that the differential eq.~(\ref{e:mf}) has the following solution if the magnetic field and obliquity angle do not evolve with time:
\begin{equation}
\frac{P^2(t)}{2} - \frac{P_0^2}{2} = P\dot P t  = (\kappa_0 + \kappa_1 \sin^2 \alpha) \beta B_p^2 t.
\end{equation}
In the limit of initial spin period $P_0 \ll P(t)$ we can obtain:
\begin{equation}
\tau_\mathrm{ch} = \frac{P(t)}{2\dot P} .
\end{equation}
This age is called the spin-down (characteristic) age, and is often used to estimate the true age of a radio pulsar. This age has obvious disadvantages, related to assumptions which are made to derive it: (1) no evolution of the dipole field, (2) no evolution of the obliquity angle, and (3) the initial period is much smaller than the modern period of radio pulsars. We show lines for constant spin-down ages in Figure~\ref{pdotp} together with periods and period derivatives of radio pulsars from the ATNF pulsar catalogue\footnote{https://www.atnf.csiro.au/research/pulsar/psrcat/}.

In a simplified picture, the pulsar evolution in the $P$~--~$\dot P$ plane proceeds along the lines of constant magnetic field. A pulsar is born somewhere in the upper left corner of this diagram and descends toward the death line crossing lines of constant spin-down age. After the pulsar crosses the death line it becomes radio silent. In this review we show that NSs might have very different paths. They could rise (dipolar magnetic field increases) and fall (dipolar magnetic field decays). 

One of the characteristics which is used to study magnetic field evolution is the braking index. It is computed as:
\begin{equation}
n = 2 - \frac{P\ddot P}{\dot P^2} .
\end{equation}
If the dipole field and obliquity angle do not evolve in time, it is expected that $n=3$. If the field is multipolar, with dominant component $l$, the braking index is $n=2l+1$ \citep{Krolik1991}. The braking index could therefore be a powerful tool to study both the configuration and evolution of NS magnetic fields. Unfortunately, the braking index is hard to estimate reliably from observations, due to a variety of reasons, such as difficulty in measuring $\ddot P$ and irregular behaviour of radio pulsars (so-called red noise). 

Besides dipolar poloidal magnetic fields, NSs could also have: (1) small-scale poloidal fields, (2) large- or (3) small-scale toroidal fields in the crust. The presence of small-scale poloidal fields is often discussed as an essential mechanism which increases the curvature of open field lines, and thus leads to the activation of the pulsar mechanism in long-period radio pulsars. Arcs of poloidal magnetic fields can be seen in phase-resolved X-ray spectroscopy because such arcs trap electrons and thereby form absorption lines by inverse Compton scattering. These arcs are located at small distances from the NS, and therefore form phase-dependent absorption features. However, there is also an alternative explanation that these absorption spectral features are in fact proton cyclotron lines \citep{2013Natur.500..312T}.

Toroidal magnetic fields can be present in the NS crust. This field component decays and heats the crust, increasing the total NS thermal flux\cite{Perna2013MNRAS}, which seems to be a promising mechanism to explain magnetar quiescent emission \citep{igoshev2021}. Large-scale toroidal magnetic field could also cause the magnetospheric twist in magnetars which leads to an increase in the magnetospheric currents, which is further seen in X-ray spectra because low-energy X-ray photons are scattered by these currents, which creates a power-law scaling at energies above $2-3$~keV \cite{Lyutikov2006}.

The origin of strong and structurally complicated NS magnetic fields is still a matter of considerable debate. The birth of a NS is a rather violent process. Thus, there are numerous possibilities to enhance and restructure the field at the proto-NS stage. A brief summary of possibilities as to how NS magnetic fields could arise is given in Sec.~2.

This complicated magnetic field structure results in non-trivial field evolution, and many interesting observational consequences. The main tendency of the field evolution is decay. At relatively early stages of a NS's life ($\lesssim 10^6$~yrs) it can be dependent on the NS cooling, if the star is not too massive to switch on rapid neutrino losses. Thus, field and thermal evolution are inter-related \cite{Aguilera2008A} as field decay provides heating. The latter effect is especially pronounced in magnetars, where field evolution is governed by its restructuring due to a Hall cascade. This process can end in a so-called Hall attractor --- another hypothetical stage under active study.
On longer timescales a decrease of the field depends on properties of the crust, which in turn depend on the NS mass and history of accretion. Deep crustal properties are not well-known, bringing additional serious uncertainties in calculations of magnetic field behaviour. The whole picture of field changes can become much more complicated if processes in the NS core are at least equally important in comparison with crustal ones. On top of this, early observational appearance of the field can be blurred due to an episode of significant fall-back accretion immediately after the NS formation. External fields can be diminished and re-configured by a strong flow of matter, and the field structure and magnitude would slowly return to previous values (taking into account the overall field decay) by diffusion through the accreted layer. Main processes of field decay and re-emergence are discussed in Sec.~3.

Due to many factors in play, magnetic field evolution can be very non-monotonic, which leads to the appearance of various sources corresponding to different stages of evolution. Some observational appearances of field evolution, especially those related to rapid release of magnetic energy, are well-known (but not necessarily equally well understood). Best examples can be found in magnetar observations. Other observable features are less prominent, and their nature is not so clear. We sketch out the present-day observational picture in Sec.~4.
 
The richness of potential manifestations of different aspects of NS magnetic field evolution deserves a separate discussion of potential near-future discoveries and underlying theoretical explanations. In Sec.~5 we discuss several items, including thermal maps of NS surfaces, polarisation measurements which might be available in the next few years, and signatures of the Hall attractor in statistical properties of NSs, as well as in parameters of individual sources. As we are still far from a complete understanding of magnetic field behaviour in NSs, there are many open problems. We specify and discuss them in Sec.~6.

\section{Formation}

Neutron stars are the most magnetised objects in the universe, with field strengths up to at least $10^{15}$~G. There are two leading hypotheses for the formation of such powerful magnetic fields: (1) fossil field origin and (2) dynamo at the proto-NS stage. Modest growth of the magnetic field can also be obtained due to thermoelectric effects \cite{1980SvA....24..425U, 1983MNRAS.204.1025B}, but magnetar-scale fields can hardly be reached this way. 

The key element of the fossil field hypothesis is that the radius of a NS and its progenitor --- an O or B star --- differ by many orders of magnitude. Specifically, a NS has a 10-14~km radius, while a typical O or B star has a radius $R_\mathrm{OB}\sim10R_\odot$. If magnetic flux is conserved during the supernova explosion, the field is amplified by a factor $(R_\mathrm{OB}/R_\mathrm{NS})^2 \approx 5\times 10^{11}$ \cite{1964SPhD....9..329G}. Some 7-12\% of O and B stars are strongly magnetised with field strengths reaching 10 kG, which could then easily produce NS magnetic fields up to $\approx 5\times 10^{15}$~G \citep{Igoshev2011,Makarenko2021}.

Modern MHD simulations \cite{Schneider2019Natur} showed that strong stellar magnetic fields could be formed as a result of stellar merger. Authors \cite{Schneider2019Natur} found that the star which is born as a result of a merger has magnetic field $\approx 9$~kG which is compatible with values for some strongly magnetised massive stars \cite{Makarenko2021}.

Note though that a compact object is formed only from the core of the massive progenitor star. Thus, it is incorrect to substitute the whole stellar radius in the estimate of magnetic field amplification due to flux conservation. As was shown in \cite{2008AIPC..983..391S}, at most a few percent of the total progenitor magnetic flux can be used for the formation of a NS field. This brings us to the conclusion that some additional field amplification is necessary to explain magnetar-scale fields. 

The idea of dynamo amplification of fields at the proto-NS stage was suggested some time ago  \cite{1992ApJ...392L...9D, 1993ApJ...408..194T}. Earlier analytical studies already showed that proto-NSs are unstable to convection \citep{Miralles2000ApJ,Thompson2001,Miralles2002ApJ}, because they are not transparent for neutrinos. Thus, the convection transports heat released deep inside the collapsing star towards the outer layers, which are transparent for neutrino cooling. In these earlier studies it was found that the maximum poloidal magnetic field which could be generated as a result of convection is $5\times 10^{16}$~G \cite{Miralles2002ApJ}. They also found that toroidal magnetic field could be 100-300 times stronger than the poloidal magnetic field.

More modern three-dimensional simulations of the supernova explosion \citep{Nagakura2020} confirm that strong convection develops in a proto-NS. If some magnetic field is present, the convection can increase it by many orders of magnitude and produce both poloidal and toroidal components. However, a rapid rotation (initial spin periods $\lesssim 6 $-7~msec) is required to form a strong dipolar component \cite{2020SciA....6.2732R}. If a NS does not rotate rapidly enough, it might result in a stochastic dynamo which generates strong small-scale fields with a weak dipolar component.

In some other studies on this subject the magnetic field is amplified at the cost of rotational energy, i.e.\ small initial spin periods are necessary in order to reach fields $\sim10^{15}$~G.
The inevitable prediction of ``millisecond magnetars'' can potentially be tested. On one hand, analysis of parameters of supernova remnants (SNRs) of magnetars did not show any signatures of additional energy releases due to rapid rotational energy losses \cite{2006MNRAS.370L..14V}. On the other hand, the success of models with additional energy release fitting supernova lightcurves \cite{2010ApJ...717..245K}  (and sometimes gamma-ray bursts \cite{2018ApJ...869..155S}) provides indirect support for the existence of ``millisecond magnetars'' \cite{2021arXiv210310878D}, and thus to models with field amplification at the cost of rotational energy.

Newborn NSs can pass through an episode of fall-back accretion which could bury the magnetic field deep in the crust \citep{1989ApJ...346..847C}. This fall-back accretion continues for days and months (much longer than the crust solidification time which is about $\approx 10$~s \cite{Pons1999}) with decreasing mass accretion rate up to $3\times 10^{2}$~M$_\odot$/year at the beginning. The total accreted mass can reach $0.1$~M$_\odot$, or even more. However, smaller values might be more typical. The pressure of magnetic field with strength of $\sim 10^{12}$~G is not enough to stop this accretion, so such fields could be buried deep within the crust at densities sometimes exceeding $10^{12}$~g~cm$^{-3}$.

Most often magnetic field does not decay rapidly enough in the crust, and diffuses back to the surface on timescales of $10^4$~--~$10^6$~years \citep{Ho2011,2012MNRAS.425.2487V,Bernal2013,Igoshev2016}. It occurs due to a combination of three physical effects \cite{Igoshev2016}: (1) condensed poloidal field forms strong toroidal field, (2) interaction between poloidal and toroidal fields creates additional forces which push poloidal field towards the surface (re-emergence on Hall timescale) and (3) poloidal fields generally diffuse in the crust and thus re-emerge toward the surface (re-emergence on Ohmic timescale). In the case of a supernova with strong fall-back, even magnetar-strength fields could be buried, leading to ``hidden magnetar" cases \citep{1999A&A...345..847G}.

Overall, NSs are born with a range of magnetic field strengths ($10^{11}$~--~$10^{15}$~G) and shapes (dipolar, multipolar, toroidal)  due to stochastic processes during the dynamo amplification. Some NSs (Central Compact Objects) are born with a strongly suppressed dipolar component either due to fall-back accretion or due to features of the dynamo amplification process. These stars might be expected to grow their dipolar field component with time. Other NSs might be born with stronger poloidal magnetic fields combined with even stronger ``hidden" (crust-confined) toroidal magnetic fields. This variety of initial fields should lead to a great diversity of observational appearances \citep{Kaspi2010,Igoshev2014} which we discuss in Sec.~\ref{s:observations}.

\section{Evolution (Theory)}

Magnetic fields could be present both in the NS crust and core. When the NS core cools down it turns into a superconductor \citep{Baym1969Natur}.
Researchers often assume that protons in the core form a Type I superconductor, and thus expel any magnetic field due to the Meissner effect. In this case only magnetic field evolution in the crust is modelled, e.g.\ \citep{Gourgouliatos2016,Gourgouliatos2018}. If the core is a Type II superconductor, magnetic field should be concentrated into isolated vortices. Alternatively, the core might demonstrate a mixture of these two states (1.5 Type superconductor). Which types of superconductivity could be present in a NS core is a matter of active scientific research, see e.g.\ \cite{Wood2020}.   

If a magnetic field is present in the core, its evolution differs significantly from that of fields in the crust.  A NS crust is formed of a lattice where only electrons can move freely. In contrast, the core is expected to have multiple charge carriers. Therefore the main drivers for magnetic field evolution in the crust are its finite conductivity and Hall evolution, while in the core ambipolar diffusion seems to be the essential effect.

\subsection{Crustal magnetic field}

Goldreich \& Reisenegger \citep{Goldreich1992} suggested the most commonly used framework for theoretical studies of NS magnetic field evolution. They started with one of Maxwell's equations (Faraday's law):
\begin{equation}
\frac{\partial \vec B}{\partial t} = - c \nabla \times \vec E ,
\end{equation}
where $\vec B$ and $\vec E$ are the magnetic and electric fields, respectively. They supplemented this equation with Ohm's law to derive:
\begin{equation}
\frac{\partial \vec B}{\partial t} = - c \nabla\times \left(\frac{1} {4\pi en_e}  (\nabla \times \vec B) \times \vec B + \frac{c}{4\pi \sigma} \nabla \times \vec B \right) ,
\label{s:mfd_eq}
\end{equation}
where $\sigma$ is the electrical conductivity (inversely proportional to resistivity), $n_e$ is the local electron number density, and $e$ is the elementary charge. The first term on the right side of this equation corresponds to the Hall effect, and the second to the Ohmic decay. Note also that the magnetic field which is being described here is the field within the NS itself, which cannot be directly measured. It is only the portion of the field which extends outward from the star that can be observed, and it is only the large-scale, dipolar part that affects the spin-down, as described in the Introduction.

\subsubsection{Ohmic evolution}

The partial differential eq.~(\ref{s:mfd_eq}) can be solved analytically for some simple special cases. For example, pure Ohmic decay is described as:
\begin{equation}
\frac{\partial \vec B}{\partial t} = - \frac{c^2}{4\pi \sigma} \Delta \vec B ,  
\end{equation}
where $\nabla\times \nabla \times\vec B = \Delta\vec B$. If we decompose the field as a sum of poloidal and toroidal parts $\vec B = \vec B_p + \vec B_t$, the equation for the poloidal part becomes:
\begin{equation}
\frac{\partial \vec B_p}{\partial t} = - \frac{c^2}{4\pi \sigma} \Delta \vec B_p .
\label{e:polod_ohm}
\end{equation}
(The toroidal part does not enter here since Ohmic decay occurs completely independently for poloidal and toroidal components.)
If we are interested in solutions which could be represented by the Fourier transform:
\begin{equation}
\vec B_p (t, \vec r) =\vec B_0 \exp(i\omega t - i\vec k\cdot \vec r) ,
\end{equation}
then we can simplify the differential eq.~(\ref{e:polod_ohm}) even further:
\begin{equation}
i\omega = -\frac{k^2 c^2}{4\pi \sigma} .
\end{equation}
If we are interested only in magnetic field decay with the timescale $\tau_{Ohm}= -\frac{1}{i\omega}$ for a field which has a spatial extent $L = 1/k$ then we obtain:
\begin{equation}
\tau_\mathrm{Ohm} = \frac{4\pi \sigma L^2}{c^2} .
\label{e:timescale_ohm}
\end{equation}
There are two main estimates for the spatial extent $L$. If a NS is assumed to be a uniformly conducting body filled with matter with conductivity $\sigma$, then $L = R_\mathrm{NS}/(2\pi)$. If it is instead assumed that the magnetic field is confined to the crust, then $L = h_\mathrm{crust}$, where $h_\mathrm{crust}$ is some typical scale height of the crust. Eq.\ (\ref{e:timescale_ohm}) means that the magnetic field decays exponentially on a timescale $\tau_\mathrm{Ohm}$:
\begin{equation}
\vec B_p (t, \vec r) =\vec B_0 (\vec r) \exp\left(-\frac{t}{\tau_{Ohm}}\right) .
\end{equation}

As is clear from eq.\ (\ref{e:timescale_ohm}), the crustal conductivity determines how quickly the magnetic field decays. The conductivity itself is composed of two components:  (1) conductivity due to phonons, and (2) conductivity due to crystalline impurities. The total conductivity can then be computed as:
\begin{equation}
\sigma = \frac{\sigma_\mathrm{ph}\sigma_\mathrm{Q}}
{\sigma_\mathrm{ph} +\sigma_\mathrm{Q}} ,
\end{equation}
where $\sigma_\mathrm{ph}$ and $\sigma_\mathrm{Q}$ are contributions related to phonons and impurities, respectively. Thus, for the timescales we can write $\tau_\mathrm{Ohm}^{-1}=\tau_\mathrm{Ohm, ph}^{-1}+\tau_\mathrm{Ohm, Q}^{-1}$. If we substitute numerical values for conductivities  \cite{2004ApJ...609..999C}, we can get the following estimates for these timescales. For $\sigma_{Q}$:
\begin{equation}
\sigma_{Q} = 4.4\times 10^{25} {\;\mathrm s}^{-1}\, \left( \frac{\rho_{14}^{1/3}}{Q}\right)
\left(\frac{Y_e}{0.05}\right)^{1/3}\left( \frac{Z}{30}\right) .
\label{e:conduct_q}
\end{equation} 
In this equation  $\rho_{14}$ is the density in units $10^{14}$~g~cm$^{-3}$, and $Y_e$ is the electron fraction in the current layer.
The parameter $Q$ characterises how ordered the crystalline structure of the crust is: $Q=n_\mathrm{ion}^{-1}\Sigma_i\,n_i \times(Z^2-\langle Z\rangle^2)$. Here $Z$ is the ion charge, and $n$ is number density.

The conductivity eq.\ (\ref{e:conduct_q}) translates to the timescale:
\begin{equation}
\tau_Q \approx 200 \; \mathrm{Myr}\, \left(\frac{L}{1 \;\mathrm{km}}\right)^2\left( \frac{\rho_{14}^{1/3}}{Q}\right)
\left(\frac{Y_e}{0.05}\right)^{1/3}\left( \frac{Z}{30}\right).     
\end{equation}
The parameter $L$ might be even smaller and comparable to $\approx 100$~m \citep{2004ApJ...609..999C}, which decreases the timescale by two orders of magnitude. For normal radio pulsars it is typically assumed that $Q < 1$. In statistical analysis of kinematic ages for normal radio pulsars \citep{Igoshev2019} it was found that $\tau_Q > 8$~Myr,  which could be  translated to $Q < 0.25$,  for some assumptions about $L$ and inner crust properties. For magnetars, it has been suggested that the impurity parameter $Q$ could be $ \gg 1$   \citep{2013NatPh...9..431P}. This hypothesis is motivated by the observed absence of magnetars and their supposed descendants (like the Magnificent Seven NSs, see below) with spin periods longer than $12$~s. Population synthesis studies \citep{Gullon2015MNRAS,Makarenko2021} with small values of $Q \leq 1$ predict that a large number of magnetars should exist with spin periods $P > 20$~s, contrary to observations. A lack of magnetars
with long spin periods could also occur due to a different reason. It was suggested \citep{Karageorgopoulos2019MNRAS} that in magnetars the spin-down currents do not penetrate sufficiently deep into the crust to allow efficient slowing down after the spin period reaches 15~s.

For the phonon conductivity we can get the following estimate:
\begin{equation}
\sigma_\mathrm{ph} = 1.8\times 10^{25}\; {\mathrm s}^{-1}\, \left(\frac{\rho_{14}^{7/6}}{T^2_8}\right)\left(\frac{Y_e}{0.05}\right)^{5/3} ,
\end{equation}
where $T_8$ is the temperature of the crust in units $10^8$~K. 
The timescale which corresponds to this equation is:
\begin{equation}
\tau_\mathrm{ph} \approx 80 \; \mathrm{Myr}\, \left(\frac{L}{1 \;\mathrm{km}}\right)^2 \left(\frac{\rho_{14}^{7/6}}{T^2_8}\right)\left(\frac{Y_e}{0.05}\right)^{5/3} .
\end{equation}
In this way the magnetic field decay depends strongly on crustal temperature and becomes irrelevant when the NS cools down and reaches temperatures $T_8 < 0.1$ at ages $t > 1$~Myr. Influence of magnetic field decay by phonons could explain some irregularities in the distribution of pulsars by spin-down ages \citep{IgoshevPopov2014}.





\subsubsection{Hall evolution}
Compared with Ohmic decay, the Hall evolution is much more complicated. The paper \citep{Shalybkov1997} was one of the first where the influence of the Hall term on the magnetic field evolution was analysed. Hall evolution was studied mostly numerically in multiple articles since \citep{Hollerbach2002,Hollerbach2004,ChoLazarian2009,Wareing2009PoP,Wareing2010JPP,Kojima2012,Gourgouliatos2014,Gourgouliatos2014PhRvL,Geppert2014MNRAS,Marchant2014ApJ,Bransgrove2018MNRAS,Brandenburg2020}.

Analogously to eq.~(\ref{e:timescale_ohm}) it is possible to write an estimate for the timescale of the Hall evolution:
\begin{equation}
\tau_\mathrm{Hall} = \frac{4\pi e n_eL^2}{cB(t)} ,
\end{equation}
where $B$ is the local instantaneous magnetic field, and $L$ is the typical spatial scale of the associated electric currents. Compared with linear Ohmic evolution, the applicability of this timescale is very limited. The Hall evolution is strongly nonlinear, coupling different spatial scales, and poloidal and toroidal components of the magnetic field together.

In earlier works it was noticed that eq.~(\ref{s:mfd_eq}) has some similarity to the vorticity equation \cite{Goldreich1992,Wareing2009PoP}; it was thus suggested that Hall evolution generates some kind of turbulence-like behaviour. Therefore, any large-scale magnetic field would evolve through a cascade-like process towards smaller scales. This behaviour results in  acceleration of the magnetic field decay according to eq.~(\ref{e:timescale_ohm}), as the spatial scale $L$ of electric currents decreases. That is, the Hall cascade is not a dissipative process by itself, but if the scale of currents diminishes with a characteristic time shorter than $\tau_\mathrm{Ohm}$ for an (initial) large spatial scale, then the cascade determines the rate of magnetic energy dissipation. As $B$ decreases, $\tau_\mathrm{Hall}$ becomes longer, and finally the Hall cascade becomes unimportant. The same is true for initially low fields, typical for most radio pulsars.

Detailed numerical simulations in 2D and 3D \cite{ChoLazarian2009,Wareing2009PoP,Wareing2010JPP,Brandenburg2020} showed that a cascade does indeed develop, and energy is transferred from large-scale to small-scale magnetic fields. However, the slope of the turbulent spectrum appears to be proportional to $k^{-5/2}$, instead of the $k^{-2}$ originally suggested by Goldreich and Reisenegger \cite{Goldreich1992}. Another key difference is that in Hall turbulence the large-scale field creates strong anisotropies in the small-scale fields --- unlike hydrodynamic turbulence, where small-scale eddies are simply advected by large-scale flows, but otherwise remain largely isotropic.

Since Hall evolution leads to development of a turbulent cascade, in some cases energy can also be transferred from small to large scales \citep{Wareing2009PoP,Wareing2010JPP,Brandenburg2020,Gourgouliatos2020MNRAS,Igoshev2021ApJ}. If a dynamo at the proto-NS stage failed to generate a strong dipolar or quadrupolar magnetic field, but generated fields with dominant $l=10-50$ (turbulent dynamo), the Hall evolution leads to energy redistribution and at least some increase in the strength of the large-scale components.

Perhaps the most important difference between hydrodynamic and Hall turbulence is that Hall turbulence often seems to retain a strong memory of its initial condition. That is, the spectra quickly evolve to look like typical turbulent spectra, but if one examines the field structures in physical space, one finds that the evolution seems to saturate after a few Hall timescales, and the magnetic field hardly changes thereafter. This effect was noted in periodic-box geometry by \cite{Wareing2009PoP,Wareing2010JPP}, and was later also referred to as `frozen turbulence' \cite{Gourgouliatos2018AG}.

Closely related to this feature is the idea of a Hall equilibrium --- a magnetic field configuration where the Hall term is identically equal to zero. Such equilibria were first proposed by \cite{Reisenegger2007AA}, and have been extensively studied further \cite{Gourgouliatos2013MNRAS,Marchant2014ApJ,Bransgrove2018MNRAS,Kitchatinov2019}. In most of these works the field is computed directly from the requirement that the Hall term $\nabla\times \left(\frac{1} {4\pi en_e}  (\nabla \times \vec B) \times \vec B\right)=\vec 0$, in which case one must additionally consider the question of whether the solution is stable or not. In contrast, if direct time-stepping calculations seem to evolve toward `frozen turbulence' solutions, then they are necessarily evolving toward a stable Hall equilibrium. The inclusion of Ohmic decay of course means that none of these solutions are true equilibria, since they are gradually decaying in time. And since Ohmic decay affects different components differently, the Hall term is also not identically equal to zero, but is constantly transferring some energy between different components to keep the overall field configuration close to the `equilibrium' configuration, with all components decaying at almost the same rate.

The idea of a Hall equilibrium can be further extended to a Hall attractor. That is, are there equilibria so powerful that a large range of initial conditions is attracted to them, even ones that are initially quite different? In periodic-box geometry the answer is almost certainly no; all initial conditions seem to retain some memory of their particular details \cite{Wareing2009PoP,Wareing2010JPP}. However, in the astrophysically relevant spherical-shell geometry, the answer is quite possibly yes, at least for axisymmetric calculations; \cite{Gourgouliatos2014,Gourgouliatos2014PhRvL,Bransgrove2018MNRAS} found that a range of different initial conditions evolved toward similar near-equilibrium solutions. Having well-defined magnetic poles and an equator is perhaps helping the solutions to focus toward a particular equilibrium in a way that periodic-box solutions are intrinsically incapable of doing. How robust such an axisymmetric attractor is to fully 3D perturbations is less clear though. If the initial condition is still predominantly axisymmetric, but non-axisymmetric structures are also allowed to develop, then \cite{Wood2015} found that the axisymmetric part of the solution is still broadly similar to the previously known attractor. On the other hand, for initial conditions that are fully 3D, there seems to be no tendency to evolve toward an attractor; the solutions seem to be more of the `frozen turbulence' type that always retain details of their particular initial condition \cite{igoshev2021}.

Returning to the question of whether Hall equilibria are stable or not, considerable work has focused on many different aspects. The papers \cite{Rheinhardt2002,Rheinhardt2004} demonstrated that many field configurations could be unstable, and suggested that this was an efficient mechanism to transfer energy from large scales directly to small scales, without requiring a turbulent cascade. How clearly such instabilities could be isolated in cases where a cascade is also present is an open question \cite{Wareing2009,Pons2010}, and probably depends on the details of any given configuration. One scenario where such a Hall instability can be clearly identified as separate from a cascade is if the initial condition is predominantly axisymmetric, and one is interested in the question of how non-axisymmetric structures develop. This case was considered by \cite{Gourgouliatos2019}, who found that while the axisymmetric component evolved following cascade/attractor theory, the non-axisymmetric structures developed according to the Hall instability theory.

Further work in Hall instabilities has also considered more local aspects, such as instabilities that can develop in small-scale current sheets \cite{Wood2014,GourgouliatosKondic,Gourgouliatos2016MNRAS,Kitchatinov2017,Kitchatinov2021}. The depth-dependence of the electron number density $n_e$ plays an important role in many of these instability modes. The nonlinear development of some of these modes also suggests that the energies released in the form of Ohmic heating can be enough to power magnetar activity.

\subsection{Core magnetic field: ambipolar diffusion}
Ambipolar diffusion was suggested by \cite{Goldreich1992} as the primary mechanism driving the magnetic field evolution in the NS core. The core is in a liquid state, so electrons, protons and neutrons can move differently. The core composition might also include more exotic particles like muons and hyperons which contribute to ambipolar diffusion. An important factor in the core is the electron-proton density. When electrons and protons move (following the electric current) into a region with different neutron density, they could experience inverse beta-decay and form neutrons, which are uncharged and thus not sensitive to electric fields. In a real neutron star this simple picture becomes much more complicated because the matter in the core is expected to be in a superfluid and superconducting state.

Until recently ambipolar diffusion was studied by making simple estimates for local velocities in the NS core, see e.g.~\cite{Gusakov2020MNRAS}. The first two-dimensional simulations suggested that ambipolar diffusion in the absence of superfluidity and superconductivity is not important for magnetar evolution \citep{2021CoPhC.26508001V}. However, newer analysis suggests that the situation is considerably more complicated. 

In the general case the equation which describes the ambipolar diffusion depends on velocity $\vec v_\mathrm{amb}$, which in turn is a function of magnetic field and pressure gradient \citep{Glampedakis2011,Passamonti2017}:
\begin{equation}
\frac{\vec f_\mathrm{mag}}{n_c} - \nabla (\Delta \mu) = \frac{1}{x_n^2} \frac{m^*_p\vec v_\mathrm{amb}}{\tau_\mathrm{pn}} ,
\label{e:v_amb}
\end{equation}
where $\Delta \mu = \mu_p + \mu_e - \mu_n$ is the deviation from $\beta$-equilibrium, $n_c$ is the local number density of charged particles such as electrons, $x_n$ is the neutron fraction, $m_p^*$ is the effective proton mass, and $\tau_\mathrm{pn}$ is the proton-neutron collision time. The magnetic force is written as:
\begin{equation}
\vec f_\mathrm{mag} = \frac{\vec j\times\vec B}{c} ,
\end{equation}
where the electric current density is 
\begin{equation}
\vec j = \frac{c}{4\pi} \nabla\times\vec B .
\end{equation}
Overall, the ambipolar diffusion velocity is related to magnetic field evolution as:
\begin{equation}
\frac{\partial\vec B}{\partial t} = -\frac{c^2}{4\pi \sigma} \nabla \times (\nabla \times\vec B) +  \nabla \times (\vec v_\mathrm{amb} \times\vec B) ,
\label{e:amb}
\end{equation}
where the first term is again simply Ohmic decay, and could be negligible for a superconducting core. If the $\beta$-equilibrium is established sufficiently quickly, then eq.~(\ref{e:v_amb}) can be simplified and a direct relation between $v_\mathrm{amb}$ and $f_\mathrm{mag}$ can be found.

Ambipolar diffusion is hard to study numerically and analytically for several reasons: (1) eq.~(\ref{e:amb}) is highly nonlinear, (2) effects of superfluidity and superconductivity, and (3) the exact composition of the core is not known yet. On the numerical side the ambipolar diffusion term is proportional to $\nabla \times [((\nabla \times\vec B) \times\vec B) \times\vec B]$, and is thus highly sensitive to the precise magnetic field configuration. At the boundary between core and crust, the magnetic field evolution could form discontinuities \citep{2021CoPhC.26508001V} which need to be treated in a sophisticated manner. 
If no $\beta$-equilibrium is assumed, it is necessary to separately solve eq.~(\ref{e:v_amb}) simultaneously with the magnetic induction equation. Effects of superconductivity organise the magnetic field into flux tubes. The interaction of these flux tubes with particles is poorly understood. For example, Ruderman \cite{Ruderman2010NewA}  suggested a model to compute the magnetic field evolution due to interaction between flux tubes and vortices when the NS spins up or slows down. This model predicts braking indices $0 < n < 0.5$ for the youngest radio pulsars and $n\approx -1$ for older radio pulsars. These predictions seem to contradict modern measurements of braking indices, see Secs.~\ref{s:growing_field} and \ref{s:isolated_pulsars}.

Overall, ambipolar diffusion leads to non-trivial magnetic field evolution. Recent estimates \citep{Passamonti2017,Passamonti2017MNRASB} on $v_\mathrm{amb}$ give values ranging from few~km/Myr for standard matter up to $10^6$~--~$10^8$~km/Myr for superfluid and superconducting cores with fast neutrino cooling. These velocities translate into shortest magnetic field evolution timescales ranging from $1$~Myr (normal matter) to $10$~Kyr (superfluid/superconducting matter in a small range of temperatures). 


\subsection{Numerical efforts}

The partial differential equations outlined in the previous section are challenging to solve numerically for multiple reasons: (1) equations are strongly nonlinear with varying coefficients which depend on local conditions, (2) equations have to be solved in spherical geometry, and (3) physics of the NS core is not understood completely. There are two basic approaches to solve these differential equations: (1) finite difference methods and (2) pseudo-spectral methods.

Finite difference methods are reviewed in \citep{2019LRCA....5....3P}, and have mostly been used for axisymmetric calculations \citep{Aguilera2008AA,Aguilera2008A,Pons2009A,Vigano2013MNRAS}, including detailed simulations of the core physics (see e.g.~\cite{Passamonti2017}). 2D finite difference codes are relatively straightforward to implement, but can be very difficult to extend to 3D, due to the coordinate singularities at the poles. An alternative is to embed the entire star in a 3D Cartesian grid \cite{Vigano2019}, but then of course the desired spherical boundaries become somewhat awkward.

In contrast, pseudo-spectral codes can be more difficult to program in some regards (e.g.\ requiring efficient forward and backward Fourier and Legendre transforms), but using spherical harmonics in the angular coordinates effectively eliminates the coordinate singularities at the poles, since spherical harmonics are the natural expansion functions in spherical geometry, and are completely isotropic in the angular coordinates. A fully 3D crustal field code \citep{Wood2015,2020ApJ...903...40D,2021ApJ...914..118D,igoshev2021} was developed by modifying a code \citep{Dormy1998,Aubert2008} originally used in the planetary dynamo context, where high-resolution pseudo-spectral codes have been in use for over two decades (e.g.\ \cite{Glatzmaier2002}). Spectral method was also used in 2D axisymmetric calculations \citep{Hollerbach2002,Hollerbach2004,PonsGeppert2007A}.

\section{Observations}
\label{s:observations}

Magnetic field evolution cannot be observed directly yet, as astronomical measurements are not numerous and precise enough. Moreover, the computed poloidal dipolar magnetic field $B_p$ depends on the same quantities as characteristic age $\tau_\mathrm{ch}$. Therefore, these two quantities cannot be used together. This is why this problem is so different from the NS cooling where usage of $\tau_\mathrm{ch}$ as an age estimate is very common. 

However, multiple data pointing towards magnetic field evolution exist. In some cases observational results seem to be robust, in others alternative interpretations are still possible. Evidence in favour of field decay comes from observations of energy release (bursts or continuous activity), spin parameters, or properties of surface emission. In this section we discuss several types of objects (including both isolated and binary NSs) and present existing arguments in favour of magnetic field evolution coming from observations of these sources.

\subsection{Magnetars}

We start with magnetars, since for this class of NSs present-day arguments supporting magnetic field decay seem to be the most compelling. The main properties of these objects are described in several recent reviews \cite{2015RPPh...78k6901T, 2017ARA&A..55..261K}. Here we very briefly summarise those features that are directly related to the subject of our review.

In the first place, magnetars are young NSs whose activity is due to energy release of strong electric currents supporting their magnetic field (note that sometimes magnetars are also defined just as NSs with very strong magnetic fields irrespective of magnetic energy dissipation, see Sec.~\ref{hmxb}). This energy output can happen in a transient manner --- we then observe short magnetar flares (see an old classical review \cite{2006csxs.book..547W} with description of the main types of magnetar bursts), or energy can be released more or less gradually in the crust (quiescent emission), which results in additional heating of a NS surface up to several million Kelvin --- a factor of a few times larger than the hottest NSs of the same age without additional heating (see Sec.~7 in \cite{2019LRCA....5....3P}). These two types of activity seem to be linked with each other, which is visible during so-called outbursts --- prolonged periods of enhanced activity \cite{2018MNRAS.474..961C}.  In these periods the surface temperature increases as well as the rate of flares.

Both types of activity require significant energy budgets. It cannot be provided by rotational energy losses, thermal energy, or accretion. Thus, it might be due to magnetic field dissipation --- the only remaining source of energy. In addition, this type of energy can easily be liberated in the form of strong bursts of electromagnetic radiation on a very short timescale, which is important to explain rapid (millisecond) development of magnetar flares with luminosities up to $\sim 10^{47}$~erg~s$^{-1}$.

Magnetic fields can provide roughly $E_{mag}\sim 2\times 10^{47}\, B_{15}^2$~erg, where $B_{15}$ is the field strength in units $10^{15}~G$. Strong magnetar external fields $B\sim10^{14}$~--~$10^{15}$~G are well-established using various techniques, including proton cyclotron line measurements, e.g.~\cite{2013Natur.500..312T}. Such an energy reservoir can power magnetar activity for several tens of thousands of years, at least. On the other hand, it is not sufficient to support continuous activity beyond a timescale of several hundred thousand years.

\begin{figure}[H]
\includegraphics[width=10.5 cm]{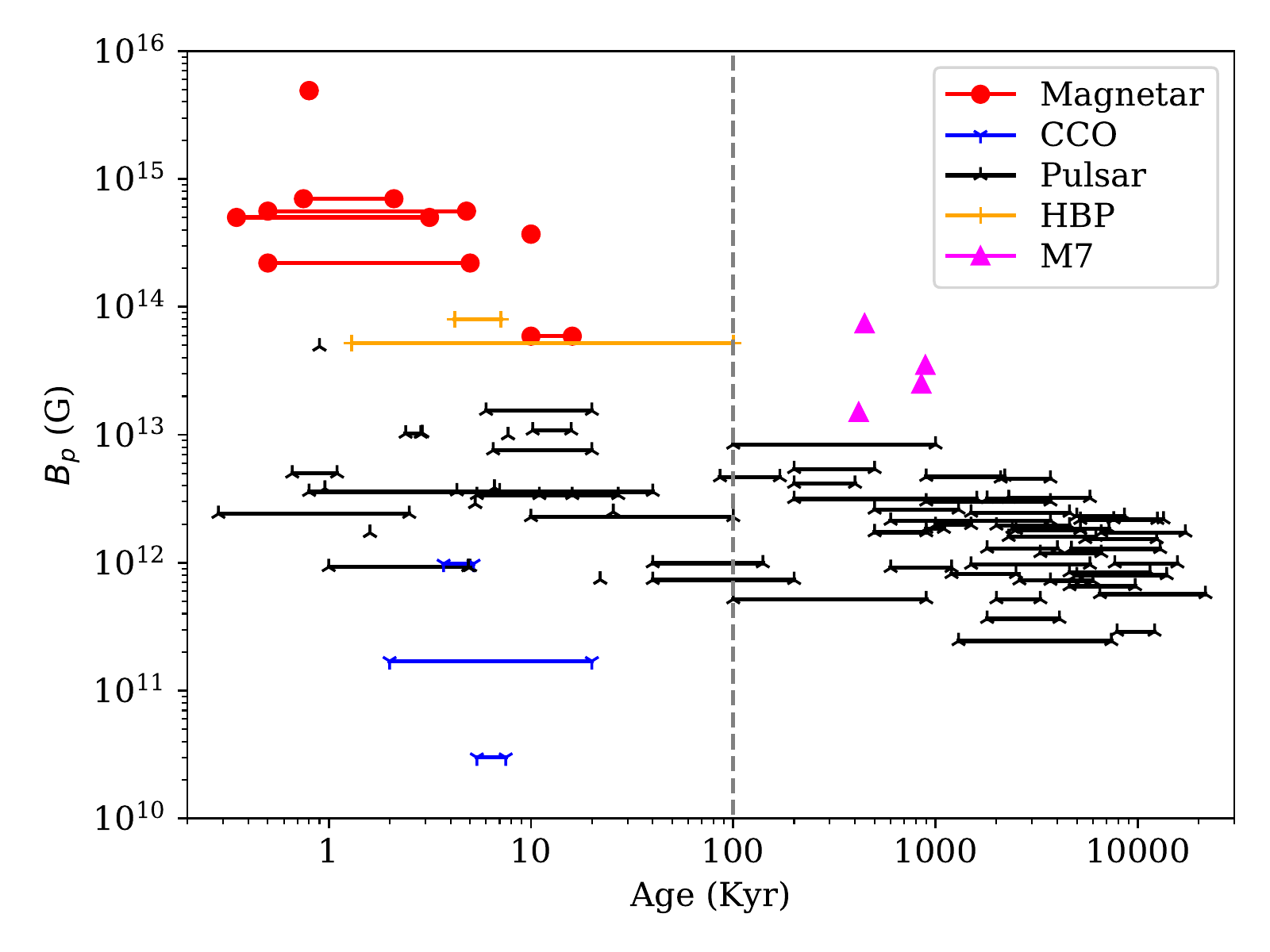}
\caption{Poloidal dipolar magnetic fields measured using period and period derivative for different types of neutron stars as a function of supernova remnant age or kinematic age. HBP are high-B pulsars, CCO are central compact objects. The length of the horizontal line shows the range of acceptable ages. Information is from multiple articles \cite{SafiHarb2017JPhCS,2012ASPC..466..191P,Rogers2016MNRAS,2018MNRAS.474..961C,2014A&A...563A..50P,Igoshev2019}. The grey dashed line separates regions where age is determined as an age of the associated supernova remnant (left part) and where the age is determined based on NS kinematics (right part). }
\label{f:age}
\end{figure}

Known magnetars are young objects, see e.g.\ Figure~\ref{f:age}. This can be established via their associations with SNRs and young stellar clusters, their galactic distribution, and, finally, via spin parameters. The so-called characteristic age, $\tau_{ch}=P/(2\dot P)$, for these sources is typically $\sim10^3$~--~$10^4$~yrs (except cases of so-called low-field magnetars, sources with strong small-scale fields, but standard --- $10^{12}$~--~$10^{13}$~G --- dipole fields, see a review in \cite{2013IJMPD..2230024T}). These age estimates are in good correspondence with the magnetic energy budget and its rate of release, supporting the standard picture where magnetar activity is due to field decay.

Episodes of magnetar activity are often accompanied by strong and rapid variations in spin properties. The onset of an outburst can result in significant changes in $\dot P$, as in the case of 1E 1048.1-5937 \cite{2020ApJ...889..160A}. At moments of strong bursts anti-glitches (i.e., a momentary slow-down of the NS rotation) are detected \cite{2016MNRAS.458.2088P}. This indicates that the magnetosphere is significantly distorted during these periods of enhanced energy release. This is expected if activity is initiated by reconfiguration of magnetic field in the crust of a NS; see e.g.~\cite{2018MNRAS.481.5331A, 2019MNRAS.484L.124C} and references therein.  

As they age, magnetars are expected to become less and less active \cite{2011ApJ...727L..51P}. Finally, we might obtain a NS with spin period of a few seconds or a little more, slightly hotter than expected for a radio pulsar of the same age (and mass) and, probably, with a different configuration and strength of the magnetic field.

\subsection{Magnificent Seven}

The Magnificent Seven (M7) also known as X-ray Dim Isolated Neutron Stars (XDINS) are nearby cooling isolated NSs (see a review \cite{2009ASSL..357..141T}). They are characterised by dominance of thermal surface emission. Magnetic fields of these seven sources are mainly estimated either from $P$, $\dot P$ measurements, or from spectral features interpreted as proton cyclotron lines (see a table e.g.\ in \cite{2014A&A...563A..50P}).  All field values are in the range $10^{13}$~--$10^{14}$~G, i.e.\ just below those of magnetars. No bursts (or any other type of significant transient activity) have ever been detected from any of the M7 objects. This points to a significantly lower rate of magnetic energy release and, probably, to a more relaxed field structure.

The M7 objects are considered as possible descendants of magnetars, e.g.~\cite{PonsGeppert2007A,2010MNRAS.401.2675P}. Spin periods of the M7 NSs are compatible with practically vertical tracks going down from the magnetar region in the $P$--$\dot P$ diagram in correspondence with the scenario of rapid decay of strong magnetic field on a timescale $\lesssim$~few hundred thousand years.

In general, absence of magnetars and M7-like sources with $P\gtrsim20$~s is indirect evidence of field decay, since with an initial surface field $B_0\sim10^{14}$~G a NS could spin down to a period of a few hundred seconds in a few hundred thousand years (a typical age of the M7 sources) if the field were not decaying. Recent examples of population studies of magneto-rotational evolution of initially highly magnetised NSs can be found in  \cite{2019MNRAS.487.1426B, Makarenko2021}. These studies demonstrate that rapid field decay in magnetars prevents formation of bright sources with spin periods longer than those already observed for these objects. Such evolution inevitably brings aged magnetars in the region of the $P$~--~$\dot P$ diagram where the M7 sources are situated. To simultaneously reproduce $P$, $\dot P$, kinematic ages, and thermal properties, M7 NSs might experience significant field decay.

Because of their proximity, distances and proper motions are measured for several of the M7 NSs with good precision. This allows to measure their velocities and trace back trajectories, until the place of origin (e.g.\ an OB association) is reached. Such analyses provide an estimate of the age of the NS. Kinematic ages of these NSs are typically smaller than characteristic ages \cite{2014A&A...563A..50P}. The latter are obtained under the assumption of constant magnetic field. If the field is decaying, then the characteristic age is an upper limit. We can thus conclude that in the case of the M7, the magnetic fields decayed during their evolution.

Relatively large surface temperatures of the M7 NSs deserve additional comments in the context of discussion of their field evolution. 
Even low-mass solutions, which provide slower cooling due to absence of direct URCA processes \cite{2015SSRv..191..239P}, cannot explain  thermal luminosities of the M7 sources for their estimated ages \cite{2020MNRAS.496.5052P}. This supports the idea presented by \cite{2007PhRvL..98g1101P} (see also \cite{2008A&A...483..223U}) that surface temperatures of magnetars, M7, and maybe some other highly magnetised NSs, are determined not just by residual heat, but by magnetic energy released in the process of field decay. In the case of the M7, temperature enhancement is not as large as for magnetars, but is still needed to bring all the data into correspondence.

\subsection{Central Compact Objects}

Another peculiar type of NSs with relatively high surface thermal emission are central compact objects in supernova remnants (CCOs in SNRs); see a list and brief review in \cite{2021A&A...651A..40M}. They have temperatures $\lesssim2 \times 10^6$~K for typical ages of a few thousand years \cite{2020MNRAS.496.5052P}, see Figure~\ref{f:age}. In several cases the high temperatures of such objects can be explained by standard cooling without direct URCA processes and with light element accreted envelopes \cite{2004ARA&A..42..169Y}. However, in some cases additional information does not allow such a solution.

The lightcurve of the CCO in SNR Kes 79 demonstrates a high pulsed fraction of surface X-ray emission. Such a feature can be explained only if the high temperature part of the surface has a very small area. Otherwise, due to light-bending the pulsed fraction could not be that large. The area of hot spots depends on heat release and transport in the crust, both of which depend on the magnetic field. Large crustal fields can allow formation of small hot areas on the surface due to its influence on thermal conductivity and/or due to additional energy release by the field decay. This might result in a large pulsed fraction. Exactly this was demonstrated for Kes 79 in \cite{2012ApJ...748..148S}. The authors reproduced observed properties of the NS under the assumption that the crustal field has a value in the magnetar range. Thus, the CCO in Kes 79 is one of the so-called ``hidden'' magnetars \cite{1999A&A...345..847G}.

The hypothesis of hidden magnetars is based on the idea that strong fall-back accretion \cite{1989ApJ...346..847C} just after a supernova explosion can result in submergence of magnetic field \cite{1995ApJ...440L..77M}. During the subsequent evolution the field slowly re-emerges due to diffusion through the outer plasma layer \cite{Ho2011, 2012MNRAS.425.2487V}.

A peculiar example of a ``not-so-hidden magnetar'' is given by the CCO in SNR RCW103. This NS was discovered by the {\it Einstein} observatory \cite{1980ApJ...239L.107T}. Later on it was found that the source is not only strongly variable on a timescale of years, but also has a very long --- 6.7 hours --- period \cite{2006Sci...313..814D}, which is now assumed to be the spin period of the NS, the longest known so far for an isolated compact object. It was proposed that the NS is a hidden magnetar, and long-term flux variations are due to energy release in its crust \cite{2015PASA...32...18P}. However, later it was discovered that this source demonstrates magnetar-type bursts (see \cite{2019A&A...626A..19E} and references therein for earlier observations of X-ray flares). Thus, the object represents a specific type of young magnetar, which probably passed through an episode of fall-back with disk formation which resulted in a significant spin-down (see \cite{2021ApJ...917....2X} and references therein for different studies of the role of fall-back disks in spinning down a young NS).

Obviously, significant fall-back with a typical value $\Delta M\sim (10^{-5}-10^{-3})\, M_\odot$ (or maybe more in some cases) can also bury a normal magnetic field $B\sim(10^{11}-10^{13})$~G \cite{1995ApJ...440L..77M}. Some CCOs can be such sources. Later on, the magnetic field might slowly diffuse back out on a timescale $\sim 10^3-10^4$~yrs \cite{Ho2011}. Thus, we have a different kind of evolution in comparison with magnetars and M7: the external magnetic field is increasing. Of course, this process might be accompanied by usual field decay in the crust, but the net result for standard fields on the diffusion timescale is growth of the external dipole component (see Sec.~\ref{reemerge}). This might explain absence of matured CCOs, distinguishable by their relatively short spin periods, among radio-quiet cooling NSs like the M7 \cite{2012ASPC..466..191P}. As the external field grows, a radio pulsar can start to operate. Such a source will then move up in the $P$--$\dot P$ diagram. A few pulsars with such behaviour exist \cite{2011ApJ...741L..13E, 2015MNRAS.452..845H}. Their nature is not entirely clear, but emerging field is one of the options.

\subsection{Radio pulsars with seemingly growing dipolar magnetic field} 
\label{s:growing_field}

If the spin evolution of a NS is described by eq.~(1), then one can define the braking index, eq.~(7). Spin-down of a radio pulsar is mainly determined by the dipolar field, as it decreases less rapidly with distance from the star than other multipoles do.  In the simplified standard picture decaying magnetic field might result in $n>3$, and growing field in $n<3$. Of course, the real picture is somewhat more complicated (e.g., glitches might spoil a smooth picture of spin evolution). Still, in some cases $n\ne 3$ might indicate field decay or growth (see an illustrative figure in \cite{2015MNRAS.452..845H}). In this subsection we briefly discuss a few cases of radio pulsars with presumably growing magnetic field as indicated by their braking indices.

According to the ATNF catalogue \cite{2005AJ....129.1993M} there are nine pulsars with braking indices in the range $0<n<3$. For four of them, including the Crab pulsar, the value is not far from the classical value $n=3$. For the remaining five though, the index is $\lesssim 2$, i.e.\ $\dot P$ is definitely growing. In \cite{2013IAUS..291..195E} the author lists six more pulsars with $n\lesssim 2$. Among them, PSR J0537-6910 was shown to have a quite complicated spin evolution with average $n$ between glitches about 7 \cite{2018ApJ...864..137A}, but the long-term average braking index is small. Qualitatively, the same is true for PSR B1800-21, PSR B1823-13, and the Vela pulsar. For others (as well as for a few more sources) the situation is less clear due to numerous glitches, see  \cite{2017MNRAS.466..147E}.

One well-studied example of a NS with a probably growing external field is PSR 1734-3333, with $n\sim 0.9$ \cite{2011ApJ...741L..13E}. This source demonstrated monotonic movement in the $P$--$\dot P$ diagram for $\sim14$~yrs. If this process continues at least another $\lesssim 30$ kyr, this NS will reach the magnetar region of the diagram. In the standard approach this would be interpreted as continuous growth of the dipolar field, maybe due to field re-emergence after fall-back accretion. This might be the best candidate among normal radio pulsars for a NS with increasing magnetic field.

Another candidate is PSR B0540-69. This pulsar demonstrates a very complicated behaviour: after a long period of stable behaviour \cite{2015ApJ...812...95F} it changed its braking index from $\sim 2.1$ down to 0.03, and then in a few years rose up to 0.9 \cite{2019NatAs...3.1122G}. One interpretation of such evolution was via field modification due to instabilities in the NS interior \cite{2020MNRAS.494.1865W}.

\subsection{High-mass X-ray Binaries}
\label{hmxb}

High-mass X-ray binaries (HMXBs) are systems where the primary star is more massive than $\sim 8$~M$_\odot$, and the secondary is a NS or a black hole. In this review we are interested only in systems where the secondary star is a NS. The class of HMXBs can be divided into several subclasses; in most of them a donor is either a Be-star or a supergiant. Be/X-ray binaries are identified by the presence of H$\alpha$ emission \citep{Reig2011}. In the case of supergiant binaries, the massive star is evolved and left the main sequence already. In both cases the mass transferred onto a NS interacts with a rotating magnetic field; this is the primary reason why we observe unusual X-ray activity.
 
\begin{figure}[H]
\includegraphics[width=10.5 cm]{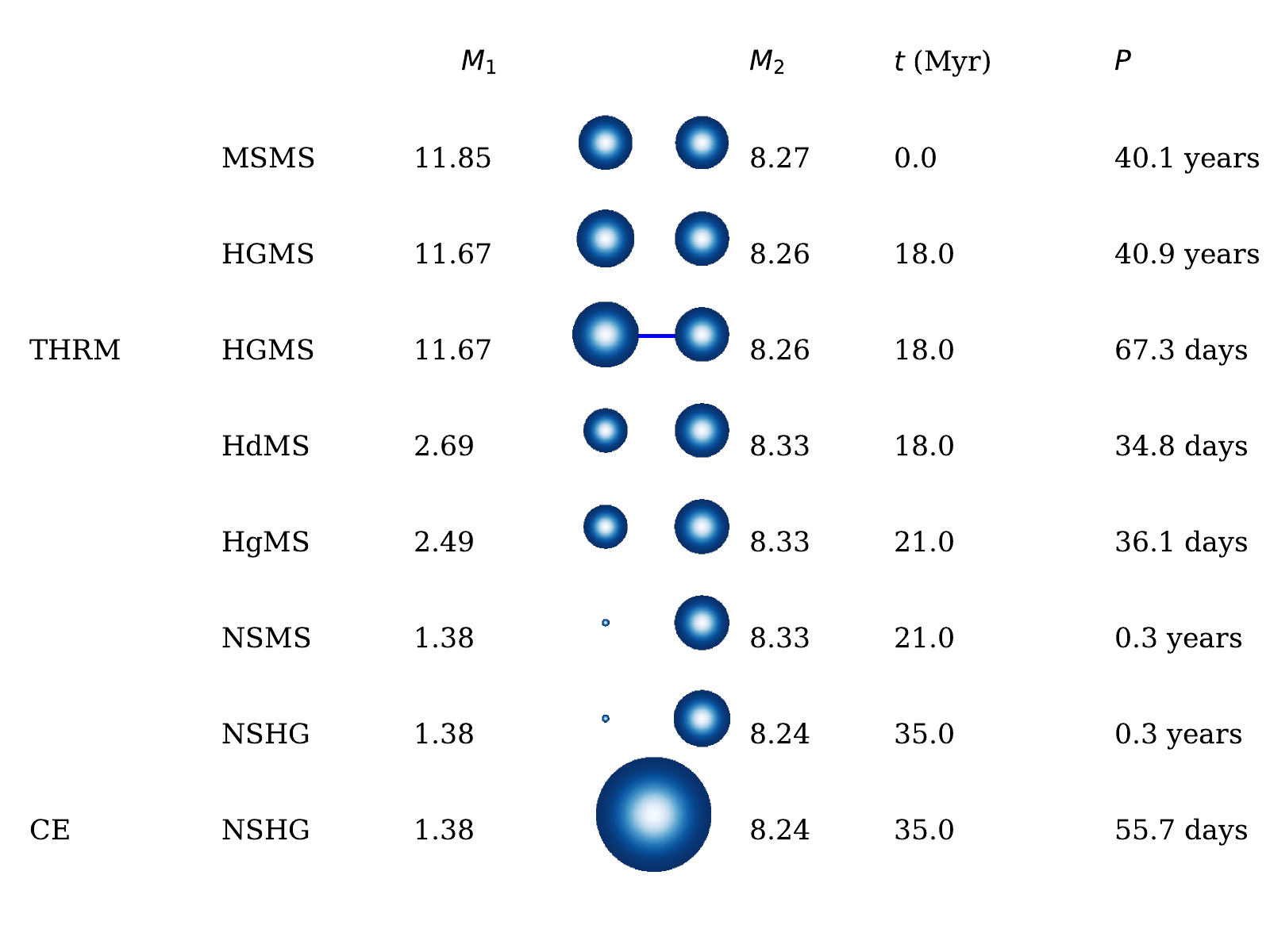}
\caption{The formation channel for Be/X-ray binaries. $M_1$ and $M_2$ are masses of the primary and secondary stars, respectively, in solar masses, $t$ is the age of the system and $P$ is the orbital period. The sizes of each circle correspond to stellar radius. THRM stands for mass transfer on thermal timescale, and CE stands for common envelope. MSMS stands for two main sequence stars, HG is a helium giant, hb is core helium burning star, Sg is AGB star, HG is Hertzsprung gap, sg is sub-giant, NS is neutron star, WD is white dwarf. \label{BeX}}
\end{figure}

The formation of Be/X-ray binaries is a complicated process which involves a non-conservative mass transfer prior to a supernova explosion \citep{Zwart1995,Shao2014,Vinciguerra2020}. This mass transfer spins up the secondary star\footnote{Definition of the primary and secondary star in these evolved systems is ambiguous because the initial primary (more massive) star loses a significant part of its mass due to the mass transfer and becomes a NS which is apparently less massive than the original secondary star. We follow the convention generally accepted in the binary stellar evolution community, and call the originally more massive star the primary star.} and allows a formation of the decretion disk. We illustrate the main formation channel for Be/X-ray binaries in Figure~\ref{BeX}. Due to the supernova explosion the NS orbit gains a large eccentricity and inclination. Therefore, a NS interacts with the decretion disk only occasionally close to the periastron passages.

In the case of the supergiant HMXB, the secondary star has a strong stellar wind which is captured by the gravitational field of the NS and further channelled by its magnetic field. In this case there is no preferable part of the orbit where accretion is possible as in the Be/X-ray binaries.

In both cases there are two independent approaches to measure the NS magnetic field: (1) based on NS spin properties and X-ray luminosity, and (2) based on X-ray absorption lines
(see e.g.\ \cite{2019Ap&SS.364..198Y} and references therein). The first approach usually suggests that the NS spin period stays close to the equilibrium period. This equilibrium period is determined by the balance between spin-up and spin-down torques, which roughly corresponds to the situation when the outer part of the magnetosphere rotates with the Keplerian velocity \citep{Davidson1973}. When a NS rotates  faster than the so-called 
accretion period, $P_A$ (defined by equality of the magnetospheric radius and the corotation radius), matter is mostly expelled due to the propeller effect, and the compact object rapidly spins down. Otherwise matter is accreted by the NS. This estimate depends significantly on properties of the matter and its interaction with the NS magnetosphere. Also, there are other approaches, based on maximum spin-up or spin-down of an accreting object, or on other methods, see some examples and references e.g.\ in \cite{2012NewA...17..594C}. 

The second approach probes magnetic fields much deeper in the magnetosphere, above the region where X-ray radiation is produced. In this case, a complicated picture of relation of the line energy on X-ray luminosity can be observed. E.g., in V0332+53 the energy  changes linearly with the source luminosity being accompanied by changes in pulse profile \cite{2006MNRAS.371...19T}. Oppositely, in A 0535+262 it is excluded that there is a positive correlation of the cyclotron line  energy with the accretion rate  \cite{2019MNRAS.487L..30T}.

\begin{figure}
\includegraphics[width=10.5 cm]{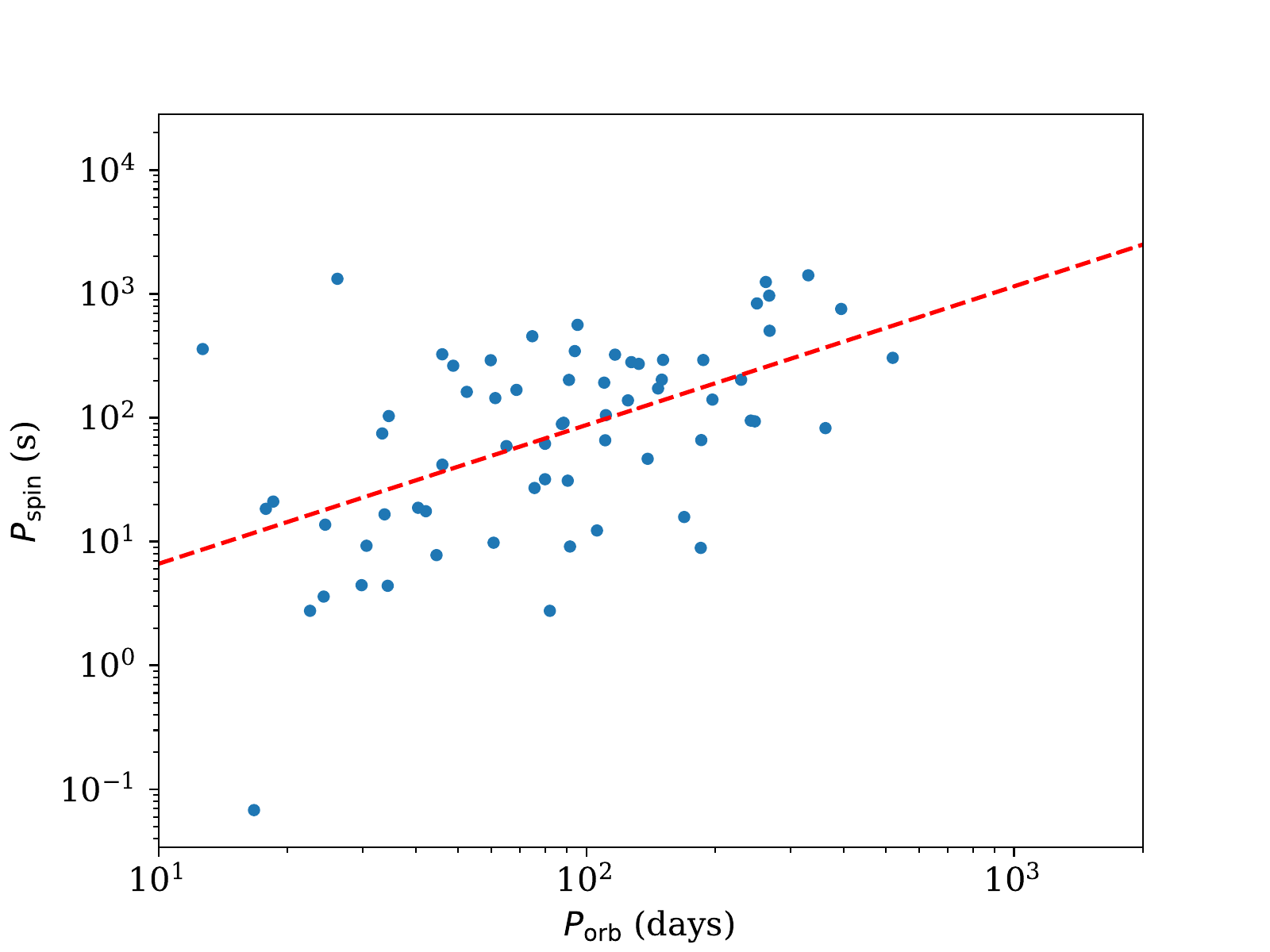}
\caption{Corbet diagram for Be/X-ray binaries. Dots correspond to systems found in the Galaxy, Small and Large Magellanic Clouds. The red dashed line shows the linear fit eq.\ (\ref{linearfit}). \label{Corbet}}
\end{figure}

We show in Figure~\ref{Corbet} the so-called Corbet diagram \citep{Corbet1986} for Be/X-ray binaries with measured orbital and spin periods in the Milky Way, Small and Large Magellanic Clouds, based on the catalogue by \cite{2005A&AT...24..151R}. The correlation seen in this figure is explained by the presence of magnetic fields. Our linear fit has the following form:
\begin{equation}
\log_{10}\left( \frac{P_\mathrm{spin}}{\mathrm{s}} \right) = 1.12 \log_{10} \left(\frac{P_\mathrm{orb}}{\mathrm{d}}\right) - 0.298 .
\label{linearfit}
\end{equation}

\begin{figure}[H]
\includegraphics[width=10.5 cm]{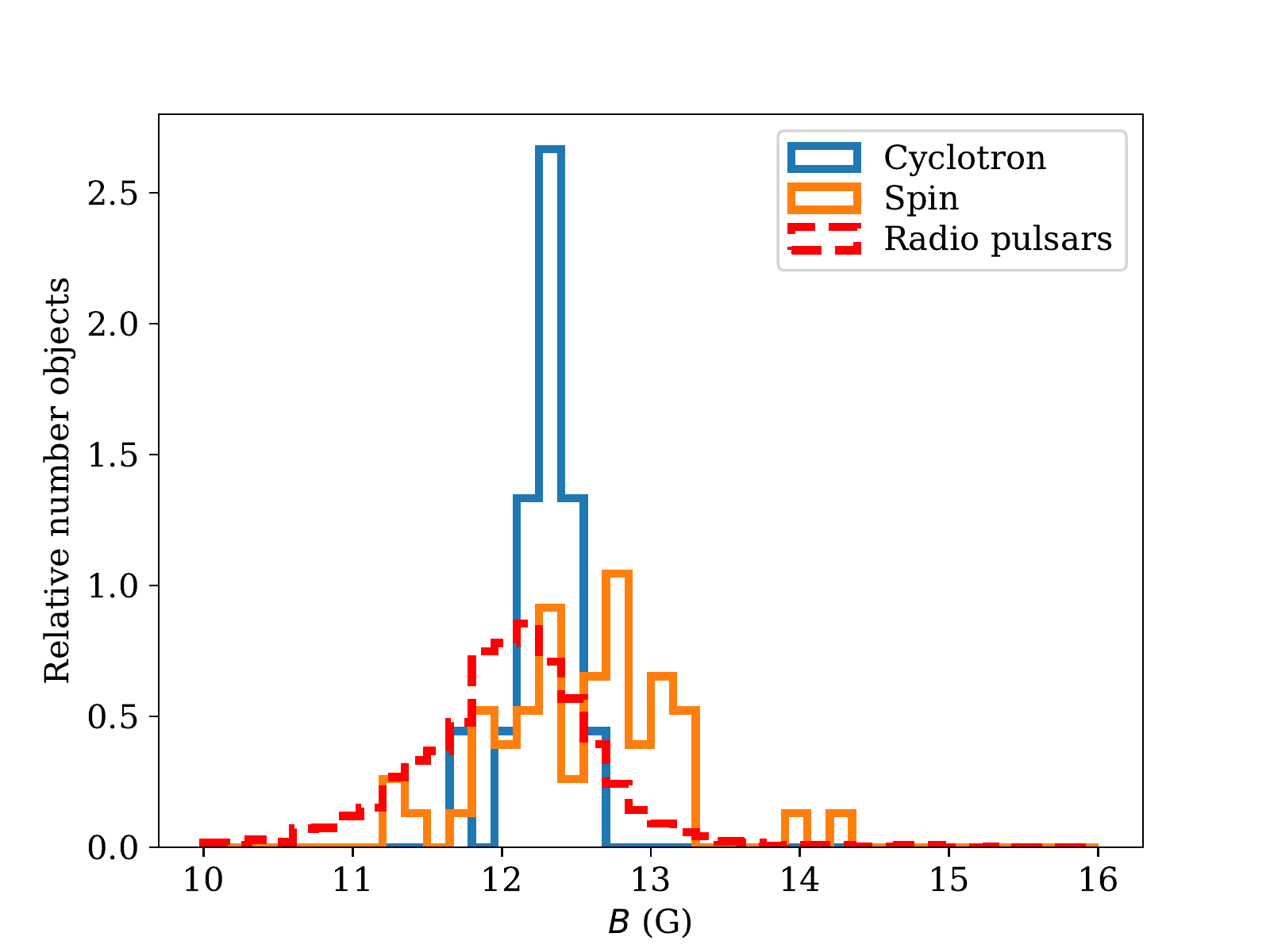}
\caption{Comparison of magnetic field measured by cyclotron resonance scattering features for accreting NSs \cite{Makishima1999} and dipolar magnetic fields for normal radio pulsars estimated using their spin period and its derivative.}
\label{cyc}
\end{figure}

The accretion period for a NS is \citep{Reig2011}:
\begin{equation}
P_\mathrm{A} \approx 23\, \mu_{30}^{6/7}\, \dot M_{15}^{-3/7}\, \left(\frac{M_\mathrm{NS}}{M_\odot}\right)^{-5/7} \; \mathrm{s} ,
\end{equation}
where $\mu_{30}$ is the magnetic moment of the NS and $M_{15}$ is the mass accretion rate in units $10^{15}$~g/s. Therefore, it is possible to estimate the NS magnetic field using its spin period, orbital period and X-ray luminosity due to accretion. For example, \cite{2012NewA...17..594C} gathered different estimates for magnetic field based on different models of accretion. Here we use their hybrid model to estimate magnetic fields for objects shown in Figure~\ref{Corbet}. We show the histogram for magnetic fields in Figure~\ref{cyc}. This analysis shows that there are two objects with magnetic fields $B_p\approx 10^{14}$~G. Other methods also occasionally give estimates for some NS fields in the magnetar region (see a brief review e.g. in \cite{2018MNRAS.473.3204I}). Whether any of these stars really are magnetars is unknown. No magnetar-like activity (bursts, outbursts) have ever been detected from these objects. Also, there are reasons to suspect that magnetars seldom appear in accreting binary systems \cite{2006MNRAS.367..732P, 2016A&AT...29..183P}. Still, in many cases, especially for ultra-luminous X-ray sources (ULXs) showing X-ray pulsations, the hypothesis of accreting magnetars was proposed by different authors, e.g.\ \cite{2015MNRAS.448L..40E, 2015MNRAS.454.2539M} (see a recent review on ULXs, including discussion of models with magnetars-scale magnetic fields, in \cite{2021AstBu..76....6F}). Critics of the magnetar explanation of pulsating ULXs can be found in \cite{2019MNRAS.485.3588K}. We discuss possible properties of accreting magnetars in the context of NS magnetic field evolution in Sec.\ 5.4.

The estimates for magnetic fields based on cyclotron line measurements are shown in Figure~\ref{cyc}. These fields are similar to those of isolated radio pulsars, except for the strongest ones ($B_p > 5\times 10^{12}$~G) and the weakest ones ($B_p < 5\times 10^{11}$~G). Ages of Be/X-ray binaries are hard to determine because the massive star was spun up and accreted material in the past. It is also unclear if the decretion disk survives for the entire main sequence lifetime of a B star, or if the star slows down and loses its disk on a shorter timescale. The lifetime of a Be star is probably limited by $2-20\times 10^{7}$~years depending on its mass. Thus, strong magnetic fields measured for NSs in HMXBs ($B_p \sim 10^{12}$~G) indicate that there is no significant magnetic field decay on timescales up to $\approx 2\times 10^8$~years \cite{2012NewA...17..594C}. This is also in correspondence with results in \cite{2019Ap&SS.364..198Y} where the authors also concluded that fields $\lesssim 10^{13}$~G do not decay significantly.


\subsection{Millisecond Radio Pulsars}

Millisecond radio pulsars (MSPs) are among the oldest NSs known; for a review see \cite{Bhattacharya1991reviewMSP}. These radio pulsars have short spin periods ranging from a few to $\approx 30$~milliseconds. Their period derivatives ($\dot P \sim 10^{-18}$~--~$10^{-22}$) are 3-4 orders of magnitude smaller than those of normal radio pulsars \cite{Bhattacharyya2021}. Significantly smaller period derivatives are thought to be related to much smaller dipolar magnetic fields, $B\sim 10^8$~--~$10^{10}$~G. MSPs are found in binaries much more frequently than normal radio pulsars. Half of all known millisecond radio pulsars are in binaries \citep{vandenHeuvel1993ASPC}, compared with only $<5.3$\% of normal radio pulsars \citep{Antoniadis2021MNRAS}. A typical MSP companion is an evolved star such as a white dwarf or neutron star.  Thus, it is natural to think that MSPs are closely related to binary pulsars. MSPs are frequently found in globular clusters where normal radio pulsars are typically absent \cite{Ransom2008IAUS}.

It is generally thought that MSPs obtain their special properties due to stable mass transfer from a secondary star to the NS at the low-mass X-ray binary stage \cite{Alpar1982Natur}. Transferred mass has large angular momentum which spins up the NS. Some interaction between accreted material and the NS magnetic field is probably responsible for weak magnetic fields of MSPs.

In particular, \cite{Taam1986ApJ} noticed that NSs in high-mass X-ray binaries accrete approximately few$\times 10^{-3}$~M$_\odot$ and keep magnetic fields $\approx 10^{12}$~G, while NSs in low-mass X-ray binaries accrete up to $0.1-0.8$~M$_\odot$ and have magnetic fields $10^{8}$~--~$10^9$~G. It was later suggested \cite{Shibazaki1989Natur} that the magnetic field after accretion correlates with accreted mass according to the dependence:
\begin{equation}
B = \frac{B_0}{1 + \dot M t / m_B} ,
\label{e:field_mb}
\end{equation}
where $m_B$ is a constant describing properties of magnetic field decay, $\dot M$ is the mass accretion rate, $t$ is the accretion time and $B_0$ is the initial magnetic field. It was found \cite{Taam1986ApJ} that eq.\ (\ref{e:field_mb}) predicts values of $B$ and $P$ which are consistent with observed or estimated values for binary and millisecond radio pulsars if $m_B \approx 10^{-4}$~M$_\odot$. Thus accretion of $10^{-4}$~M$_\odot$ decreases the magnetic field by a factor of two. This model is empirical, thus no physical mechanism explains this decay in detail. Another confirmation for this picture is provided by observational properties of binary radio pulsars with low-mass companions. It was shown that orbital periods of these objects correlate with their magnetic fields \cite{vandenHeuvel1995AA}. Shorter orbital periods correspond to larger accreted mass and smaller magnetic fields. 

In the literature it was also suggested \cite{Taam1986ApJ,vandenHeuvel1984Natur} that some MSPs are formed via the accretion-induced white-dwarf collapse; thus these objects are much younger and could therefore have stronger magnetic fields (around $10^{11}$~G) in comparison to MSPs.

The formation of a MSP in a binary is a long process. After the supernova explosion, the NS could operate as a normal radio pulsar for some time ($\sim 10^7$~--~$10^8$~years), while the secondary star stays on the main sequence. The main sequence lifetime for the secondary in a low-mass X-ray binary could reach $0.1-3$~Gyr \cite{Tauris2012MNRAS}. After leaving the main sequence the secondary expands, fills its Roche lobe and initiates the mass transfer. This is the stage when the binary is seen as a low-mass X-ray binary. 
Multiple NSs in low-mass X-ray binaries demonstrate  thermonuclear X-ray bursts which could occur only if the magnetic field is below $10^{11}$~G \citep{vandenHeuvel1993ASPC}. 
When mass transfer ceases the secondary typically turns into a white dwarf or NS, and the primary NS starts operating as a radio pulsar again. That is why MSPs are often called recycled radio pulsars. Once a NS is recycled, it can live for Gyrs as an active radio pulsar; this is one of the reasons why MSPs are found among old globular cluster population.

Over the years researchers suggested multiple explanations for decreased magnetic field of MSPs. Their smaller magnetic fields could be explained as the result of accretion \cite{Bisnovatyi1974SvA,Zhang2006MNRAS}. Magnetic field could decay in the NS crust due to its decreased conductivity while the crust is heated by accretion \cite{Geppert1994MNRAS}. It was also suggested \cite{Romani1990Natur} that magnetic field could be screened by accreted material, but similarly to CCOs the magnetic field could re-emerge, in this case on short timescales (less than 1~Myr). Very recently \cite{Cruces2019MNRAS} the role of ambipolar diffusion in magnetic field evolution was highlighted. The researchers suggested that ambipolar diffusion leads to significant magnetic field decay \textit{before} the accretion is initiated. The heating during the accretion, and rotochemical evolution after, stabilise the decayed field. Overall, all these hypotheses are quite schematic, and detailed understanding of magnetic field evolution in old NSs with and without accretion episodes is required.

The high-energy emission of MSPs is produced in their magnetosphere or due to interaction between magnetospheric currents and the NS surface. For example, gamma-rays are produced entirely in the magnetosphere \cite{Harding2021arXiv}. The thermal X-ray emission is produced due to the heating of the NS surface by return currents. The bulk soft X-ray emission is typically absent or corresponds to very low temperatures $T\leq 10^5$~K \cite{2002A&A...387..993P, 2009ASSL..357.....B,Becker1999AA,Zavlin2007ApSS}.

Until very recently the soft X-ray radiation of MSPs was attributed to emission of hot polar caps coinciding with poles of dipolar magnetic field. This assumption was based on the old ages of typical MSPs. A few years ago, the NICER X-ray telescope on board the International Space Station performed long observations of selected MSPs with well-constrained mass to probe the mass-radius relation and equation of state. Detailed modelling of surface thermal patterns \cite{Riley2019,Miller2019ApJ} showed that the soft X-ray emission is produced in two (or three) separate hot regions which do not coincide with the dipolar magnetic poles. One of these regions is elongated. The most natural explanation for formation of such regions is a complicated configuration of NS magnetic field \cite{Bilous2019}. Thus small-scale magnetic fields could survive or be formed on timescales of Gyrs.

Recent numerical simulations \cite{Suvorov2020MNRAS} showed that a magnetic field configuration somewhat similar to observations of PSR J0030-0451 can be formed as a result of accretion if the accretion flow is equatorially asymmetric. However, the authors used small accreted mass few$\times 10^{-5}$~M$_\odot$, which is significantly less than typically accreted by MSPs. Moreover, the authors did not model the core field and effects of heat on the crustal magnetic field evolution.


\section{Theory versus Observations}

 In this section we discuss several promising approaches to probe different aspects of magnetic field evolution in NSs. At the moment, none of them provide clear signatures. In some cases analyses already performed nevertheless demonstrate interesting results; in others we can expect significant progress in the very near future thanks to new observational instruments.

\subsection{Magnetic fields of isolated radio pulsars}
\label{s:isolated_pulsars}

In a lowest-order approximation, the magnetic fields of standard isolated radio pulsars might evolve mainly due to the Ohmic decay caused by impurities. The timescale of this process is expected to be longer than $\sim 10^7$~years due to several reasons: (1) we observe a large number of isolated radio pulsars, and their formation rate even without magnetic field decay is comparable to 
the supernova rate in the Galaxy \citep{Keane2008}, (2) we observe radio pulsars which move towards the Galactic disk from large distances, which means that their magnetic fields survive on timescales comparable to the vertical oscillation period in the Galaxy ($\sim 100$~Myr)\footnote{assuming that pulsars are born predominantly in the youngest part of the thin Galactic disk}, e.g.\ \cite{Noutsos2013}, and (3) there are strongly magnetised NSs ($B\sim 10^{12}$~--~$10^{13}$~G) in HMXBs which have ages $\sim 10^7$~years; see \cite{2012NewA...17..594C} and references therein. 

Because the magnetic field decay timescale is probably so long, researchers use statistical techniques to study the field decay in normal radio pulsars, including: (1) population synthesis and (2) pulsar current. The population synthesis is a Monte Carlo technique originally developed to study binary stellar evolution; for a review see e.g.\ \citep{Popov2007}. In this technique the initial properties of NSs are drawn from fixed distributions and the rotational and radio luminosity evolution is prescribed. At the last step of the population synthesis a special function selects synthetic NSs which could be detected in a pulsar survey based on radio flux. This synthetic population is finally compared with the actually observed population of radio pulsars using information about $P$, $\dot P$, radio luminosity, and spatial distribution of NSs.

The population synthesis of radio pulsars often gives contradictory results. Some population synthesis modelling found that a very modest magnetic field decay with a very long timescale up to $10^8$~--~$10^9$~years is necessary to explain the data, or even that the data can be fitted without any field decay at all, see e.g.\ \citep{Faucher2006, 2010MNRAS.401.2675P}. On the other hand, some recent studies found extremely short decay timescales clearly incompatible with multiple other observational evidence, such as \citep{Marek2020} who found a decay timescale of $\approx 4.27$~Myr. The population synthesis of isolated radio pulsars produces a pile-up of old radio pulsars near the death line unless something is assumed about the decay of radio luminosity toward the death line or magnetic field decay.

The pulsar current technique \citep{Phinney1981} is a simplified population synthesis where the Galactic NS birthrate is assumed to be constant. Therefore the number of isolated radio pulsars within certain similar spin-down age intervals is considered as a measure of the actual age \citep{IgoshevPopov2014}. It is clear from that work that pulsar current works only until $\approx 1$~Myr, and cannot measure the long-term magnetic field decay due to decay of radio luminosity or visibility of old radio pulsars. On the other hand, within its applicability range the pulsar current shows some moderate field decay (timescale $\approx 4\times 10^5$~years) which is potentially related to short-term Ohmic decay due to phonon resistivity \citep{Igoshev2015}.

Numerical integration of pulsar trajectories in the Galactic gravitational potential is another method which could indicate how long an individual pulsar already exists. Noutsos et al.~\citep{Noutsos2013} integrated orbits of some radio pulsars with measured distances and proper motions back in time using pulsar spin-down ages. In some cases they integrated up to 0.5~Gyr. Their oldest kinematic age estimate is $2\times 10^8$~years, which is fundamentally related to the fact that this is the timescale on which a pulsar is guaranteed to cross the Galactic disk, which is assumed to be the pulsar's formation place.

A somewhat similar technique was employed by Igoshev \cite{Igoshev2019}, who analysed  positions of radio pulsars above the Galactic disk for objects with well-measured parallaxes and proper motions. He compared the kinematic and spin-down ages and found that the magnetic field decay timescale is greater than $8$~Myr, with a slight preference for a decay timescale of $\approx 12$~Myr. These results are also compatible with no decay at all.

Until recently, all reliable measurements of braking indices for isolated radio pulsars were close to the canonical value of $n\approx 3$. More modern measurements take into account instrumental effects, and reveal that long-term (timescale of tens of years) braking indices could cover the range $\sim$3-1000 \citep{Parthasarathy2019,Parthasarathy2020}. Similar and even much larger braking indices were measured earlier \citep{Hobbs2004} but with less degree of certainty and for much older radio pulsars. These earlier measurements included approximately similar numbers of positive and negative braking indices. The symmetry prompted oscillatory-like explanations \citep{Biryukov2012}, possibly related to precession. New measurements \citep{Parthasarathy2019,Parthasarathy2020} have a statistically significant lack of negative braking indices, so they disfavour any oscillations and might point to magnetic field evolution.


\subsection{Thermal maps of NSs and their relation to magnetic fields and evolution}

Hot and cold spots on a NS surface could be created by any one or more of the following effects: (1) temperature of NS deeper layers channelled by magnetic fields, (2) surface heating by inverse currents in the magnetosphere, (3) heating by interaction of accreting material from a stellar companion, and (4) rotochemical reactions. These processes are summarised for example in the recent review \citep{Harding2013}.
Different mechanisms are typical for NSs at different stages of their evolution. Thus, for magnetars it is important how heat is generated in the crust and transferred toward the surface \cite{Aguilera2008A,Aguilera2008AA}, possibly in combination with some heating from magnetospheric currents. For MSPs, the hot spots are solely formed by inverse magnetospheric currents. Below we briefly discuss all these mechanisms.

\begin{figure}[H]
\includegraphics[width=10.5 cm]{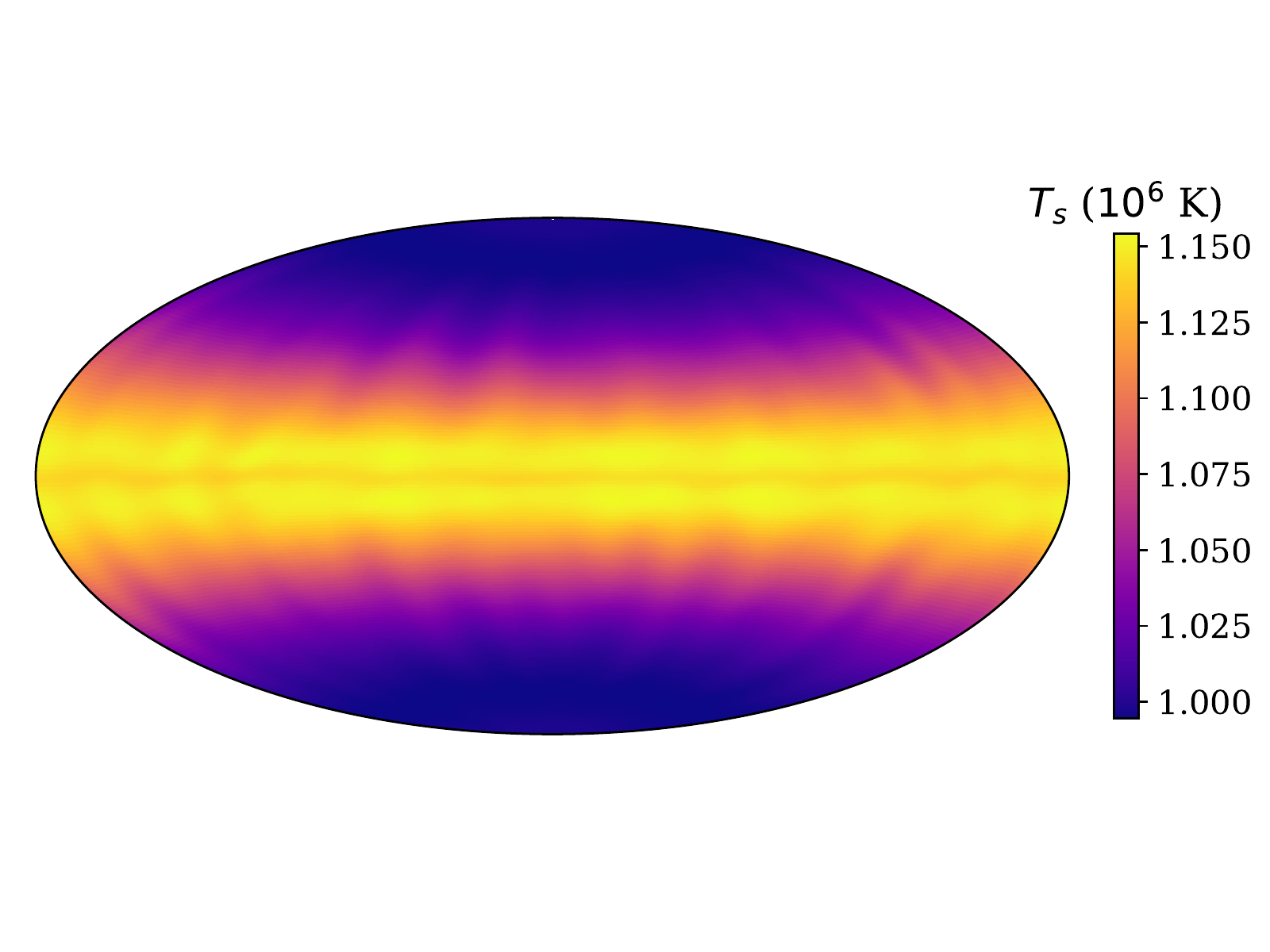}
\caption{Surface temperature distribution at age 6~kyr for model C \citep{igoshev2021} for a NS with initial poloidal dipolar magnetic field.\label{t_surf}}
\end{figure}

\begin{figure}[H]
\includegraphics[width=10.5 cm]{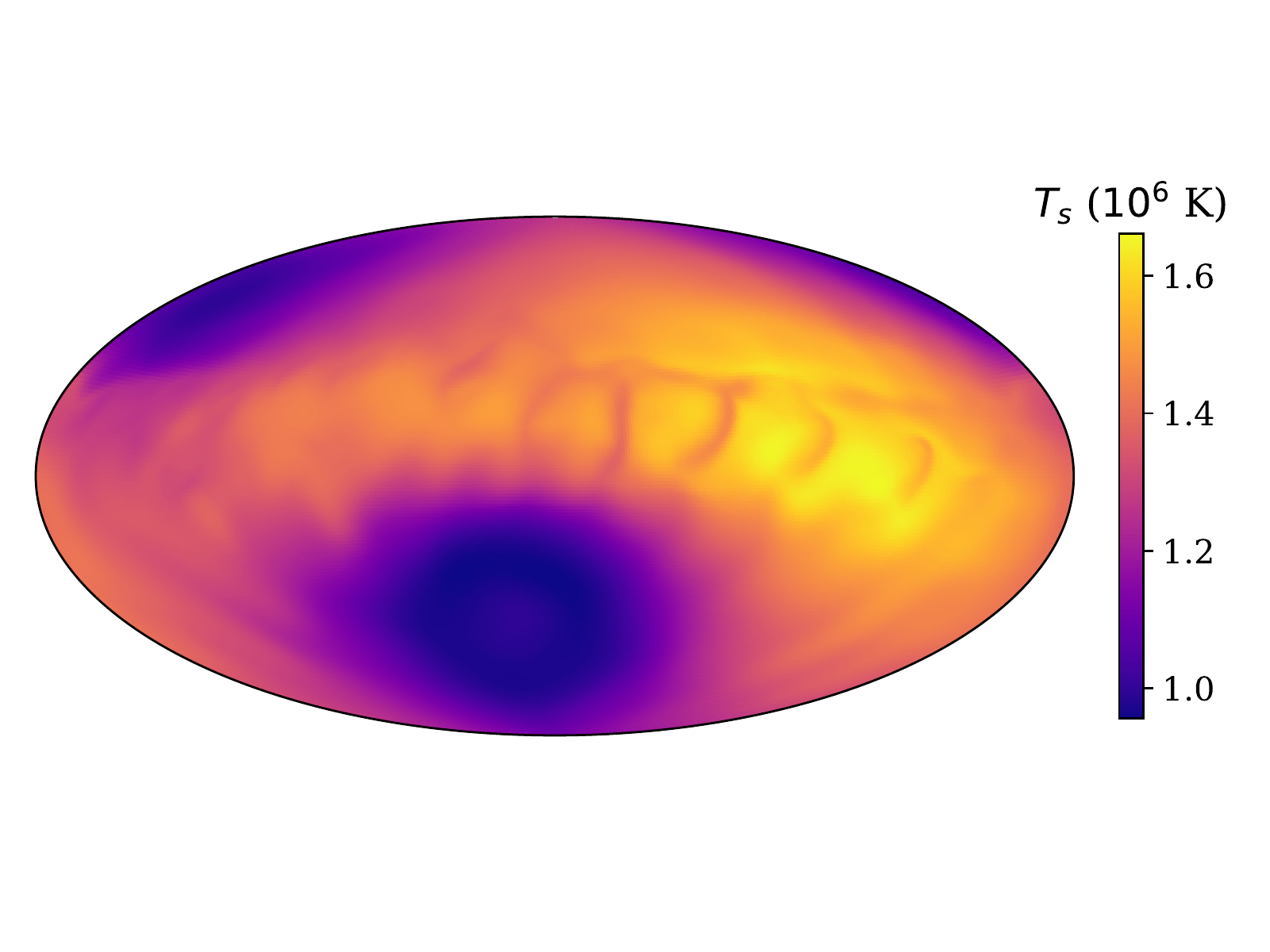}
\caption{Surface temperature distribution at age 6~kyr for model B \citep{igoshev2021} for a NS with initial poloidal dipolar field misaligned with a toroidal component by an angle $45^\circ$.\label{t_surf_B90}}
\end{figure}

The thermal conductivity of a NS crust depends significantly on magnetic field strength and orientation \citep{2015SSRv..191..239P}. Heat is transferred through the crust primarily by electrons. The electrons cannot move in directions orthogonal to the magnetic field if that field is sufficiently strong. Therefore, some regions of the NS surface become thermally connected to the core, and other regions are isolated from the core. This mechanism becomes increasingly important for magnetic fields exceeding $10^{12}$~G \citep{Geppert2004A} in the case of crust-confined magnetic field. Thus thermal and magnetic field evolution of neutron stars are coupled \cite{Aguilera2008A,Pons2009A}. If the NS field has a simple shape, e.g.\ a dipolar configuration, then the surface thermal pattern is axisymmetric and symmetric around the equator \citep{Vigano2013MNRAS,Perna2013MNRAS,igoshev2021}. Presence of toroidal magnetic field in deep layers of the NS makes the thermal pattern more complicated. If a large-scale toroidal magnetic field is present with energy comparable or larger than the energy of poloidal magnetic field, it increases the X-ray luminosity and breaks the symmetry around the equator \cite{PonsPerna2011ApJ,Perna2013MNRAS}. We show these global thermal maps in Figures~\ref{t_surf} and \ref{t_surf_B90} \citep{igoshev2021}. These regular magnetic fields form hot and cold regions with sizes ranging from a few km to NS radius size. 

\begin{figure}[H]
\includegraphics[width=10.5 cm]{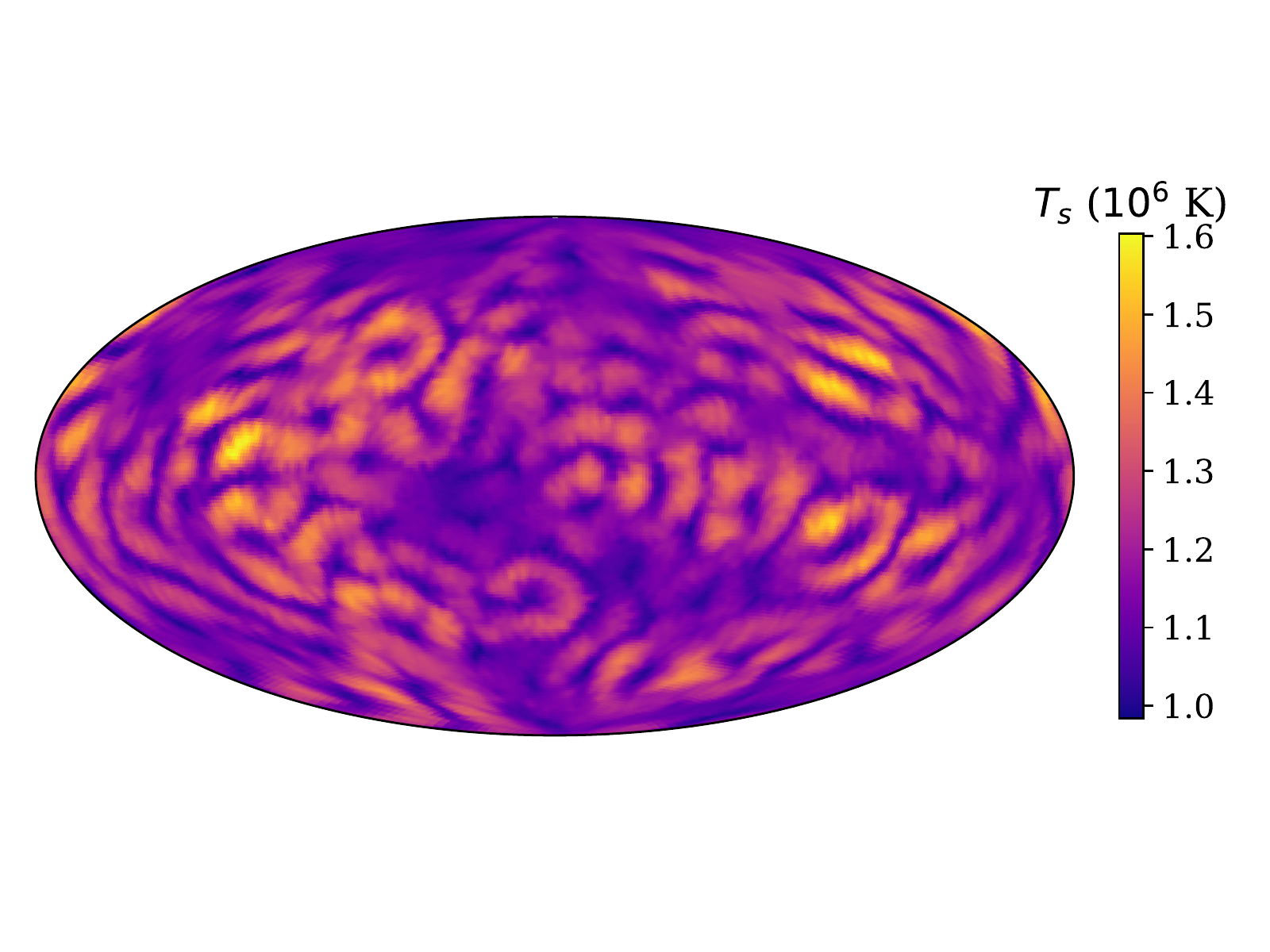}
\caption{Surface temperature distribution at age 6~kyr for model 4 \citep{Igoshev2021ApJ} for a NS with dominant small-scale poloidal field. \label{CCO_review}}
\end{figure}

In the case when small-scale multipoles dominate the magnetic field, the hot regions becomes much smaller (comparable to the linear size of the multipoles), but a large number of these hot spots are seen simultaneously \citep{Igoshev2021ApJ}, see Figure~\ref{CCO_review}.

When integrated over the visible hemisphere, taking into account the light-bending in general relativity, these thermal maps become the basis to construct lightcurves which can be further tested against timing observations. In the case of magnetars, the emission produced at the top of the atmosphere can be further modified due to inverse Compton scattering in the magnetosphere. Additional minima could thus be produced in the final lightcurves, which will depend on the state of the magnetosphere.

Conversion of observed lightcurves into positions and sizes of hot and cold regions at the NS surface is an ill-defined mathematical problem, because multiple different configurations of hot and cold regions could produce similar lightcurves. Nevertheless, this inverse problem is often attempted in the literature \citep{Hu2019}. To make this problem tractable, it is often assumed that there is only a small number of hot regions with simple shapes.

NSs with strong magnetic field and/or rapid rotation typically have a magnetosphere with strong currents \citep{Philippov2015,Philippov2018}. In the case of magnetars these currents could additionally be amplified through magnetospheric twist, which is possibly caused by crustal toroidal magnetic field or rearrangement of the crust \citep{Beloborodov2009}. Some of these currents flow through the NS surface and deposit additional heat in the region around magnetic poles.

Inverse currents (currents flowing into the NS surface) from the magnetosphere are responsible for polar cap heating in normal and millisecond radio pulsars \citep{Harding2001,Harding2002}. In this case the structure of the magnetic field affects the thermal pattern because it controls the magnetospheric currents. Thus the presence of small-scale magnetic fields forms a much smaller polar cap, or a polar cap in the shape of an annulus or crescent. Recent observations with NICER \citep{Riley2019} discovered that these thermal spots have complicated shapes in the case of millisecond radio pulsars. These complicated shapes suggest that the underlying magnetic field contains small-scale magnetic fields \citep{Bilous2019,Sznajder2020} or presence of a quadrupole component \citep{Kalapotharakos2021}.

The heating by inverse currents dominates X-ray spectra of old normal radio pulsars. For these objects, the contribution of non-thermal emission produced in the magnetosphere and the bulk surface temperature decreases significantly, and temperatures of polar caps can be measured. It is interesting to note that in some cases the sizes of these hot polar caps are much smaller than expected for a simple dipole magnetic field \citep{Zavlin2004,Igoshev2018ApJ}. In the case of millisecond radio pulsars, the bulk temperature is sufficiently low because of their age.

\subsection{Polarisation}

To study evolution of magnetic fields we need to measure present-day field values and structure as precisely as possible. This is a highly non-trivial task. Some data about field topology can be obtained by modelling of pulse profiles of magnetars; in rare cases information is derived from phase-resolved spectroscopy  \cite{2013Natur.500..312T}.  Still, new sources of information are welcomed. Polarisation measurements can become an effective new tool to probe structure of NS magnetospheres, especially in the case of strong fields.

Large magnetic fields produce quantum-electrodynamical effects which influence polarisation of emission propagating through the magnetosphere \cite{2000MNRAS.311..555H, 2002PhRvD..66b3002H}. Surface emission of NSs might also be polarised. This combination results in a  complicated picture where parameters of polarisation depend on the atmosphere of the NS (or on existence of the condensate), value of the field, its topology, and orientation of spin and field relative to the line of sight. A detailed review on all these ingredients can be found in  \cite{2019ASSL..460..301C}.

Up to now polarisation is measured for several rotation-powered radio pulsars \cite{2018Galax...6...36M}. In this case polarisation mainly reflects intrinsic properties of magnetospheric emission. For our tasks we are mainly interested in polarisation of thermal surface emission which passed through the magnetosphere. This topic is a subject of intensive theoretical studies during the last few years (see e.g.\ \cite{2015MNRAS.454.3254T} as an example of first models oriented to near-future X-ray observations) as several specialised X-ray missions have been proposed by main space agencies, and two of them --- NASA's IXPE and China's eXTP --- are approved \cite{2021ExA...tmp...62S}. These satellites will be especially effective in magnetar observations.

Complex modern models, like the one presented in \cite{2020MNRAS.492.5057T}, can include details of the surface emission (different atmospheric composition and surface temperature distribution), emission propagation (light-bending), degree of the initial polarisation, different combinations of angles related to orientation of the emitting system with respect to the observer, and finally, many effects related to the magnetic field including twisted topology, global or local.
Some of the ingredients can be determined from analysis of pulse profiles, spectral data, etc. Then, X-ray observations of degree of polarisation and polarisation angle might provide information about the magnetic field.

Polarisation of thermal emission was measured only in the optical range for one of the M7 sources --- RX J1856.5-3754 \cite{2017MNRAS.465..492M}. This is the brightest optical source among the Seven. Future generations of large telescopes will enable measurements of polarisation from other NSs of this type and also from magnetars. However, it is important to have measurements at different ranges, as modelling shows that the effect is energy-dependent. Multi-wavelength observations might thus help to disentangle different contributions to polarisation data and determine field parameters.


\subsection{Hall attractor}

The appearance of the Hall attractor stage could manifest itself in stagnation of the field decay and a specific field configuration in the crust and outside of a NS. The first manifestation can be visible either in spin evolution, or simply in conservation of a relatively large field value. The second one might be visible in thermal maps and/or in polarisation data. In this subsection we discuss three approaches already used to probe the existence of the Hall attractor stage in different types of NSs. Some of these results were already briefly summarised in \cite{2017JPhCS.932a2048P}.

In \cite{IgoshevPopov2014} the authors used the modified pulsar current approach to demonstrate that data on mid-age pulsars ($\sim$ few $10^5$~yrs) can be better interpreted by allowance of moderate field decay by a factor $\sim 2$ (see also Sec.~5.1 above). The nature of this field decay was uncertain. Also, it was not clear what switches off this decay (if field decrease is not terminated, then values rapidly become too small to be in correspondence with observations).

One possibility to explain field decay and its termination is related to the Hall cascade and Hall attractor. This option was analysed in \cite{Igoshev2015}. It was shown that a more natural explanation of results in \cite{IgoshevPopov2014} can be achieved if the dominant role in the decay is played by resistivity due to electron scattering off phonons. This type of decay rapidly vanishes when a critical temperature of the crust is reached. The timescale of the Hall cascade is too long in normal radio pulsars, and the Hall attractor stage might be reached at larger ages ($\sim 3$ initial Hall timescales) than those analysed in \cite{IgoshevPopov2014, Igoshev2015}.

A Hall attractor might be more pronounced in NSs with large initial magnetic fields. In particular, this stage can result in existence of NSs with relatively high fields at ages $\sim$ few $10^6$~--~$10^7$~yrs depending on further evolution. Such a possibility can be used to explain so-called accreting magnetars \cite{2018MNRAS.473.3204I}. 
These are X-ray binary systems in which the presence of NSs with magnetar-scale fields is suspected.

Combination of realistic values of the impurity parameter $Q$, initial field, and Hall cascade timescale allowed \cite{2018MNRAS.473.3204I} to fit parameters of two proposed accreting magnetars: ULX M82 X-2 and 4U 0114+65. Other sources were not discussed, but the general conclusion stated the possibility to explain properties of this type of objects under some realistic assumptions.

It is necessary to note that the status of all proposed accreting magnetars is still uncertain. Estimates of magnetic fields in these sources are very model-dependent. NSs in these binaries might be on average significantly older than known isolated magnetars, so co-existence of these two types of magnetars is possible only if there are significant differences between them (for example, in the parameter $Q$). Also, it is unclear why magnetars in binaries do not demonstrate any type of activity. It is worth noting that all known soft gamma-ray repeaters and anomalous X-ray pulsars are isolated objects, which deserves special attention taking into account the significant number of these sources and expected fraction of binary NSs. This feature can be related to evolutionary channels which result in magnetar formation and simultaneously prevent existence of a companion in most of cases \cite{2016A&AT...29..183P}.

The numerical modelling of crustal field evolution which suggested the existence of a Hall attractor predicted that at this stage the magnetic field has a very specific structure \cite{Gourgouliatos2014,Gourgouliatos2014PhRvL}. When the field structure is stabilised, its poloidal part  is mostly formed by dipole and octupole components. This might have a direct influence on the surface thermal distribution as heat transport in the crust (realised by electrons) is strongly influenced by the field. Thus, if we can derive a thermal map of the surface from observations, then we can compare it with predictions based on the field structure associated with the Hall attractor. This approach was pioneered by 
\cite{2017MNRAS.464.4390P} 
and then further developed by
\cite{2020ApJ...903...40D, 2021ApJ...914..118D}. 

M7 are the best targets to compare observational data with model predictions, as they are nearby and do not demonstrate non-thermal emission. Among these seven NSs it is possible to select sources which better fit the corresponding requirements. On one hand, we need bright sources for which long sets of observations with {\it Chandra} and {\it XMM-Newton} exist. On the other hand, we want to determine which NSs are closer to the Hall attractor stage.  The largest sets of observational data exist for RX J1856.5-3754 and RX J0740.4-3125. Among these two RX J1856 seems more evolved, as it is slightly colder, has lower luminosity and magnetic field; also RX J0740 demonstrates mild variability. Taking together, it is possible to make a hypothesis that RX J1856 might be at the attractor stage with larger probability. That is why, in  \cite{2017MNRAS.464.4390P} and \cite{2021ApJ...914..118D} the authors made comparison with this source.

A simplified modelling performed in \cite{2017MNRAS.464.4390P} did not demonstrate similarities between observed properties and those predicted under the assumption of the Hall attractor stage. However, a more sophisticated model developed in \cite{2021ApJ...914..118D} does not allow such a definite conclusion. Unfortunately, at the moment there are too many uncertainties, both on the theoretical side, and regarding observational determination of some key parameters (the angle between the line of sight and rotational axis, the angle between the magnetic dipole axis and spin axis, and so on). Still, it seems that the Hall attractor stage can fit the data on RX J1856. Further studies are clearly necessary.

Disclosure of observational evidence of the Hall attractor stage is of great interest for understanding of the NS magnetic field evolution. Hopefully, intensive observations of diversity of sources in the whole range of electromagnetic radiation can result in identification of objects presently at this stage.

\subsection{Ages of radio pulsars}
\label{reemerge}

The age of a radio pulsar is an important characteristic which is essential to constrain its magnetic and thermal evolution. There are multiple age estimates available at the moment for different objects: (1) spin-down age $\tau_\mathrm{ch}$; (2) age of associated supernova remnant $\tau_\mathrm{SNR}$ (see e.g.\ \cite{2021ApJ...914..103S} for a brief summary of different methods of SNR age estimation); (3) a group of kinematic ages $\tau_\mathrm{kin}$; (4) thermal age $\tau_\mathrm{therm}$. All these techniques have certain caveats and can be applied only to some radio pulsars.

The most commonly used estimate for the age of a pulsar is spin-down age. This estimate depends on the instantaneous strength of the dipolar magnetic field:
\begin{equation}
\tau_\mathrm{ch} = \frac{P}{2\dot P} = 10^{39}\, \frac{P^2}{2B_p^2} \; \mathrm{s} = 16\; \mathrm{Myr}\, \left(\frac{P}{1\; \mathrm{sec}}\right)^2 \left(\frac{B_p}{10^{12}\; \mathrm{G}}\right)^{-2}.
\end{equation}
The period is an integral over time, and thus requires longer times to adapt to a new magnetic field value. So, if the dipole field grows by a factor of two, say, the spin-down age decreases by a factor of four. Therefore, if there is an alternative estimate for the NS age, the significant mismatch between different age estimates could indicate that the dipole field recently experienced significant growth or decay.


%


As for the SNR age, this approach to estimate a NS age can be precise, but can only be applied in some cases.
Sometimes even association of a NS with a SNR is questionable.
Also, supernova remnants are short-lived objects; thus it is only possible to probe ages until $\approx 100$~Kyr.
Finally, even around young NSs a SNR shell can be absent, as in the Crab-pulsar case where only a plerion is observed.  This type of nebula does not expand with time and does not allow to probe the NS age. Instead it is sensitive to pulsar energy input.

As for the kinematic age $\tau_\mathrm{kin}$, it is a mixture of multiple slightly different techniques. NSs are rapidly moving objects ($v \propto 100$~km/s) due to the natal kick \cite{Lyne1994Natur,Hobbs2005MNRAS,Verbunt2017A,Igoshev2020MNRAS}. In some cases it is possible to associate a particular NS or radio pulsar with its probable birth site such as a particular SNR, OB associations, or the youngest part of the thin Galactic disk. Angular distance divided by proper motion gives so-called kinematic age of the radio pulsar. These techniques are applicable up to $\approx 10-20$~Myr (a fraction of the vertical oscillation period in the Galactic gravitational potential). However, in some cases researchers integrate the motion of the radio pulsar in the Galactic gravitational potential, and assign a number of ages which correspond to crossing-times of the Galactic plane \citep{Noutsos2013}. In this case it is necessary to rely on spin-down age to choose one unique kinematic age.

An isolated NS cools down with time, and its temperature decreases from $\approx 4\times 10^6$~K to below $10^5$~K in a few Myr \cite{Page2004ApJS}. Therefore, based on NS temperature it is possible to obtain some estimate for age.
Unfortunately, the cooling curve is quite uncertain currently. The cooling curve is very sensitive to the precise neutrino processes which operate inside the NS.  Direct URCA processes operate efficiently only for a certain range of NS masses, and they significantly decrease NS temperature. Moreover, a presence of strong ``hidden'' fields (small-scale poloidal or toroidal) increases the temperature of NSs. Additionally, an operating radio pulsar mechanism heats the polar caps and produces high-energy radiation which could hide the bulk NS thermal emission. Moreover, rotochemical reactions could heat old NSs. Taking into account all these caveats it is still possible in some cases to determine the NS age.

Overall, for some objects three (or even four) age estimates could be available simultaneously. Let us consider here what it means if two of these estimates coincide and the third differs drastically. Three different combinations are possible: (1) $\tau_\mathrm{therm} \approx \tau_\mathrm{ch}$, (2)  $\tau_\mathrm{therm} \approx \tau_\mathrm{kin}$, and (3) $\tau_\mathrm{ch} \approx \tau_\mathrm{kin}$. In the first case, a different $\tau_\mathrm{kin}$ suggests either that the association is wrong and the NS was born in a different place, or the NS was born from a run-away star. In the second case, a different $\tau_\mathrm{ch}$ could indicate unusual magnetic field evolution such as fast decay or growth of magnetic field. In the third case, a different $\tau_\mathrm{therm}$ could indicate that there is an additional source of heat such as a rotochemical reaction, or the NS is cooling more rapidly because some neutrino process operates more efficiently.

The recent work \cite{Igoshev2019} selected a number of systems with significant mismatch between $\tau_\mathrm{kin}$ and $\tau_\mathrm{ch}$. For example PSR B0950+08 has $\tau_\mathrm{ch} \approx 17.5$~Myr, while its most probable kinematic age is $\approx 2$~Myr (the credible 95\% interval is 0.4-17 Myr), and its temperature agrees well with this age \cite{Igoshev2019}. In the case of PSR J0922+0638 $\tau_\mathrm{ch} = 0.5$~Myr and $\tau_\mathrm{kin} = 1.6$~Myr. X-ray observations show only some indications for thermal emission from the polar cap, and did not detect any bulk X-ray radiation \cite{Rigoselli2018A}. This confirms an old age of this pulsar (or alternatively it is a massive NS which cooled down quickly).
Another example is PSR J0538+2817 with $\tau_\mathrm{ch}= 6.18\times 10^5$ years and $\tau_\mathrm{therm}\approx \tau_\mathrm{kin} \approx 4\times 10^4$~years \cite{Ng2007ApJ}. Possible explanations for this mismatch are a long initial period of the pulsar similar to its current period or that dipolar magnetic field decayed four times since its birth. 


\section{List of Open Problems}

At the end of this review we list items which, in our opinion, are of significant importance to the topic of magnetic fields in NSs. Of course, this is a very subjective set of problems. Still, we believe that their solution can significantly advance our understanding of behaviour of NS magnetic fields and their role in manifestation of these sources.

The listed topics are related to the origin of magnetic field, its evolution, and appearance of NSs. Some of them are closely linked to internal properties of these compact objects. Thus, such items require better theoretical understanding of conditions inside NSs. Other questions can be answered by new observations at different wavelengths, from radio to high-energy. Finally, complicated numerical models might help to solve several problems specified below.

\begin{enumerate}
\item	How do NS magnetic fields form? What is the role of coalescence  in the generation of magnetar-strength fields? 
\item What are consequences of field evolution for normal radio pulsars?
\item How do magnetic fields of MSPs evolve (both large- and small-scale)?
\item Do we observe any object with increasing dipolar magnetic field?
\item What are values of the parameter $Q$ in different NSs at different depths, and how are they related to electric currents responsible for magnetic fields?
\item Is the Hall attractor stage reached during a NS evolution?
\item	Are there any magnetars in binaries, and in HMXBs in particular?
\item   What is the state of the NS core superconductor? What is the timescale for magnetic field evolution in the core?
\item	What is the role of ambipolar diffusion in the core? Is it significant for normal radio pulsars? For magnetars? For millisecond radio pulsars?

\end{enumerate}

We expect that in the near future --- partly due to new observational facilities in the whole range of the electromagnetic spectrum, partly due to growing computational power and progress in numerical modelling, and, finally, due to better understanding of physical properties in neutron stars --- most of these questions will be answered.  However, inevitably new questions will appear as a result of future discoveries. This is already happening, since enigmatic fast radio bursts are more or less robustly linked to magnetars (see a brief recent review in \cite{2020Natur.587...45Z} and larger one on more theoretical aspects in \cite{2021SCPMA..6449501X}). Future gravitational-wave and (possibly) neutrino observations can contribute to this list as well. The probable discovery of old isolated accreting NSs (see e.g.\ \cite{2010MNRAS.407.1090B} and references therein) might shed light on very long-term field evolution in isolated objects. Initiation of SKA observations will dramatically increase the number of known radio pulsars and, probably, some other types of neutron stars \cite{2020SPIE11445E..12M}. All this will demonstrate again the ability of neutron star astrophysics to contribute also to fundamental physics.

\vspace{6pt} 



\authorcontributions{All authors contributed to all sections, with SP concentrating more on observational aspects, and RH concentrating more on theoretical and computational aspects.}

\funding{AI and RH were supported by STFC grant no.\ ST/S000275/1.  SP was supported by the Ministry of Science and Higher Education of Russian Federation under the contract 075-15-2020-778 in the framework of the Large Scientific Projects program within the national project ``Science''. }


\conflictsofinterest{The authors declare no conflict of interest.}


\abbreviations{The following abbreviations are used in this manuscript:\\

\noindent 
\begin{tabular}{@{}ll}
NS & Neutron Star\\
M7 & Magnificent Seven \\
CCO & Central Compact Object\\
HMXB & High-mass X-ray Binary \\
SNR & Supernova Remnant \\
MSP & Millisecond Pulsar\\
ULXs & Ultra-luminous X-ray sources \\
\end{tabular}}





\reftitle{References}


\externalbibliography{yes}
\bibliography{bibl}

\end{paracol}
\end{document}